\documentclass[prl,twocolumn,floatfix,superscriptaddress,longbibliography]{revtex4-2}

\usepackage{graphicx}
\usepackage{amsmath,amsfonts,amsthm}
\usepackage{bm}
\usepackage{fp}
\usepackage{float}
\usepackage{longtable}
\usepackage{calc}
\usepackage{placeins}
\usepackage{color}
\usepackage[normalem]{ulem}
\usepackage{threeparttable} 
\usepackage[T1]{fontenc}
\usepackage{arydshln}
\usepackage{epsfig}
\usepackage{xr}

\setlongtables
\begin{document}

\title{Aperiodic defects in periodic solids}

\author{Robert H. Lavroff}
\affiliation{Department of Chemistry and Biochemistry, University of California Los Angeles, Los Angeles, California, USA}
\author{Daniel Kats}
\affiliation{Max Planck Institute for Solid State Research, Heisenbergstra{\ss}e 1, D-70569 Stuttgart, Germany}
\author{Lorenzo Maschio}
\affiliation{Dipartimento di Chimica, Universit\`a di Torino, Torino, Italy}
\author{Nikolay A. Bogdanov}
\affiliation{Max Planck Institute for Solid State Research,
  Heisenbergstra{\ss}e 1, D-70569 Stuttgart, Germany}
\author{Ali Alavi}
\affiliation{Max Planck Institute for Solid State Research,
  Heisenbergstra{\ss}e 1, D-70569 Stuttgart, Germany}
\affiliation{Yusuf Hamied Department of Chemistry, University of Cambridge, Lensfield Road, Cambridge CB2 1EW, United Kingdom}
\author{Anastassia N. Alexandrova}
\affiliation{Department of Chemistry and Biochemistry, University of California Los Angeles, Los Angeles, California, USA}
\affiliation{Department of Materials Science and Engineering, University of California Los Angeles, Los Angeles, California, USA}
\affiliation{California Nanoscience Institute (CNSI), Los Angeles, California, USA}
\author{Denis Usvyat}
 \email{denis.usvyat@hu-berlin.de}
\affiliation{Institut f{\"u}r Chemie, Humboldt-Universit{\"a}t zu Berlin, Brook-Taylor-Str. 2, D-12489 Berlin, Germany}
\date{\today}

\begin{abstract}

To date, computational methods for modeling defects (vacancies, adsorbates,
etc.) rely on periodic supercells in which the defect is far enough from its
repeated image such that they can be assumed non-interacting. Yet, the relative proximity and periodic repetition of the defect’s
images may lead to spurious, unphysical artifacts, especially if the defect is
charged and/or open-shell, causing a very slow convergence to the
thermodynamic limit (TDL). In this Letter, we introduce a ``defectless''
embedding formalism such that the embedding field is computed in a pristine,
primitive-unit-cell calculation. Subsequently, a single (i.e. ``aperiodic'')
defect, which can also be charged, is introduced inside the embedded fragment. By eliminating the need for compensating
background charges and periodicity of the defect, we circumvent all associated
unphysicalities and numerical issues, achieving a very fast
convergence to the TDL. 
Furthermore, using the toolbox of post-Hartree-Fock
methods, this scheme can be straightforwardly applied to study strongly
correlated defects, localized excited states and other problems, for
which existing periodic protocols do not provide a satisfactory
description. 

\end{abstract}

\maketitle

\section{Introduction}

Due to its accuracy relative to computational cost, Kohn-Sham density-functional theory (DFT) has become the workhorse of computational materials study. DFT is not, however, a systematically improvable method due to lack of knowledge of the exact exchange-correlation functional, making it impossible to guarantee approaching an exact description. Wavefunction-based quantum chemistry, by contrast, does systematically approach exact solutions via hierarchies of increasingly accurate methods. Hartree-Fock (HF) theory is the lowest level of all these hierarchies. Post-HF methods, however, scale rapidly with system size, making their application to extended systems without additional (sometimes drastic) approximations difficult. Nonetheless, wavefunction-based methods have become relatively commonplace for periodic systems in recent decades, whether with actual periodic techniques \cite{pisani2008,Marsman2009,Booth2013_,usvyat2015,delben2,Booth2016,McClain2017,grueneis2018,Liao2021,Neufeld2023} or the so-called method of increments \cite{stoll1992,paulus2006} and other approaches combining periodic and finite-cluster calculations \cite{Tuma2006,nolan2009,Wen2011,Chan_science,schutz2017,alessio18}. All such techniques, however, have been nearly exclusively applied for closed-shell systems and their generalization to capture multireference character (strong electron correlation) is difficult. 

If strong correlation is localized to a particular site of a crystal or surface, such as a defect, a natural approach is to introduce an embedded fragment in which a portion of the system is treated with an accurate post-HF method while the surroundings are characterized with a periodic mean field-method such as HF or DFT. Through years a variety of such embedding schemes have been developed \cite{Govind1998,Kluner,Jacob08,bygrave2012,Manby2012,Libisch2014,Sun2016,Goodpaster2014,Jacob2014,Welborn2016,Libisch2017,Chulhai2018,Fertitta2018,Zhu2019,Lacombe2020,Cui2020,Jones2020,Pham2020,Hegely,Ma2021,WachterLehn2022,Nusspickel2022}. Particularly relevant for this work are periodic Hartree-Fock embedding schemes \cite{Birkenheuer2006,Hozoi2009,de_lara-castells2011,Stoyanova2014,masur2016,usvyat18,usvyat20,schaefer21,christlmaier21,Berkelbach21}.

Thus far, however, all embedding approaches that involve periodic treatment share the same major drawback: The defect that must be present during the periodic mean-field calculation is infinitely and periodically repeated. This leads to very slow convergence to the thermodynamic limit (TDL), causing a high computational expense for the periodic supercell calculations. Finally, in periodic boundary conditions, a non-neutral unit repeating infinitely causes the Coulomb potential of the system to diverge and a compensating background charge must be used to make the overall cell neutral. The artifacts of the unphysical background charge cause in many cases a substantial unquantifiable error.

There are also simpler finite-cluster approaches, that do not utilize periodic HF or DFT calculations to generate the embedding potential, but rather rely on electrostatic description of the infinite crystal, either directly \cite{burow2009} or in the form of effective point charges \cite{Wachters1972, sousa1993, Klintenberg2000, gelle2008, sushko2010}. These approaches do allow for a high-level treatment of insulating strongly correlated systems, radical or charged defects \cite{Illas1993, deGraaf2000, Sousa2001, Pradipto2012, Domingo2012-jb, Kubas2016, Katukuri2020, chen20, Bogdanov2021, Bhattacharyya2022, Petersen2022, Nabi2023}. However, the ambiguity of the embedding parameterization, e.g. the values of the ionic charges, and sensitivity of clusters to bond cutting requires special treatment \cite{barandiaran1988, lepetit2003, swerts2008, Larsson2022} and, in principle, limits the accuracy and transferability of such schemes.

In this Letter, we present an embedded fragment formulation which alleviates
these issues by allowing movement of fragment nuclei and change of the
fragment charge {\it after} the periodic HF calculation. In this way, the
periodic mean field, serving as the embedding field for the fragment, is
calculated using the primitive cell with no defects. This pristine periodic
calculation not only saves computational time by circumventing a supercell,
but also allows for fragments to be embedded in the defect-free, mean-field
embedding potential, boosting the convergence to the TDL. This embedding scheme also includes a possibility for a charged fragment
without compensating background charges, as the fragment itself is not
periodically repeated.

\section{Theory}

We begin with a model of a crystal's fragment, defined in terms of
local orbitals, electrons, and nuclei, embedded in the mean field of the rest
of the crystal. At this point, we keep the geometry of the fragment unaltered,
which we will refer to as a frozen fragment. It corresponds to a mere
partitioning of the crystalline space into a fragment and the embedding
environment. This setup is similar to that of the embedded fragment model of
Refs. \onlinecite{masur2016,usvyat18,christlmaier21}; however, there are a few
important distinctions. Firstly, the standard fragment approach restricts the
orbital space exclusively for the post-HF treatment on top of the full
periodic HF. Here, in contrast, the treatment of the embedded fragment starts already at the HF stage. Secondly, the fragment's nuclei are not only centers for the basis orbitals, as in the standard approach, but also hold the actual nuclear charges. In this way the overall problem is formulated not just as a frozen-environment approximation for the post-HF treatment, but rather as the HF and post-HF description of a physical fragment embedded in the periodic environment.


Despite the difference in the model, without a change in the fragment geometry
the one-electron Hamiltonian $h^{\rm frozen\, frag}$ of the present embedded
fragment approach is equivalent to the standard one introduced e.g. in
Ref. \onlinecite{christlmaier21}. For simplicity we consider a closed-shell crystal
and a closed-shell fragment:
\begin{eqnarray}
h^{\rm frozen\, frag}_{\mu\nu}&=&\left<\mu\left| -{1\over 2} \nabla^2\right|\nu\right>
+ \left<\mu\left| -\sum_{K}{Z_K \over |{\bf r}-{\bf R}_K|}\right|\nu\right>\nonumber\\ 
&&
+\sum_{i\notin{\rm frag}}\left[2\left(\mu\nu|ii\right) - \left(\mu i|i\nu\right)\right]\nonumber\\
&=&F^{\rm per}_{\mu\nu}-\sum_{i\in{\rm frag}}\left[2\left(\mu\nu|ii\right)
  - \left(\mu i|i \nu\right)\right],\label{eq:frozenfrag_h}
\end{eqnarray}
where $\mu$, $\nu$, ... are some local basis functions, $i$ -- localized occupied
orbitals of the converged periodic HF solution, $F^{\rm per}$ is the periodic Fock matrix, and the electron
repulsion integrals (ERIs) are given in the chemical notation:
\begin{eqnarray}\label{eq:ERIs}
  \left(\mu\nu|\rho\sigma\right)&=&\int d{\bf r}_1d{\bf r}_2
  {\phi^*_{\mu}\left({\bf r}_1\right)
  \phi_{\nu}\left({\bf r}_1\right)\phi^*_{\rho}\left({\bf r}_2\right)
  \phi_{\sigma}\left({\bf r}_2\right)\over \left|{\bf r}_1-{\bf r}_2\right|}.
\end{eqnarray}
It is evident that the fragment's one-electron
Hamiltonian (\ref{eq:frozenfrag_h})  contains the Coulomb and exchange potential from the
electrons of the environment and from the nuclei of both
fragment an environment.

Now we introduce a defect in the fragment by removing,
adding, and/or substituting nuclei in the fragment accompanied by a
respective update of the number of electrons. We will refer to a fragment with this
modification as an aperiodic defect, because
its structure is not repeated into the environment.

To distinguish the atoms and orbitals of the aperiodic defect their indices
will be decorated with a prime, as opposed to the ``un-primed'' indices denoting
atoms and orbitals prior to the fragment structure manipulation: $K\in{\rm frag} \rightarrow K'\in{\rm frag}$ for the fragment nuclei; $\psi_{i'}$, $\psi_{j'}$, ... for the the new occupied orbitals from the
fragment SCF. Importantly, these orbitals (as well as the virtual orbitals for the post-HF treatment) must remain orthogonal to the frozen occupied orbitals of the environment $\psi_i\notin{\rm frag}$, as otherwise electron number
conservation cannot be guaranteed. We preserve this orthogonality at level of fragment's basis
functions $\mu'$, obtained by projecting the AOs of the fragment atoms from the
occupied space of the environment:
\begin{eqnarray}
  \left|\mu'\right>=(1-\sum_{i\notin{\rm frag}}\left|i\right>\left<i\right|
  )\left|\mu\right>\label{eq:new_AO}
\end{eqnarray}
The fragment's basis orbitals $\mu'$ are centered on the nuclei after the
structure modification.
Such a choice of the basis by construction guarantees the orthogonality of the
fragment orbitals to the environment.

If the geometry of the fragment has been manipulated, the potential energy operator must be
updated. Adding the corresponding correcting terms to eq. (\ref{eq:frozenfrag_h}) yields for
the one electron operator of the aperiodic defect $h^{\rm def}$:
\begin{eqnarray}
h^{\rm def}_{\mu'\nu'}&=&h^{\rm frozen\, frag}_{\mu'\nu'}
- \left< \mu'\left| -\sum_{K\in{\rm frag}}{Z_{K} \over |{\bf r}-{\bf
    R}_{K}|}\right|\nu'\right>\nonumber\\
&&+
\left< \mu'\left| -\sum_{K'\in{\rm frag}}{Z_{K'} \over |{\bf r}-{\bf
                           R}_{K'}|}\right|\nu'\right>\label{eq:defect_h}
\end{eqnarray}
The summation $\sum_{K\in{\rm frag}}$  in (\ref{eq:defect_h}) goes over the
fragment atoms of the initial structure, while the one $\sum_{K'\in{\rm
    frag}}$ over those in the new one.
The Fock matrix of the aperiodic defect is naturally defined as:
\begin{eqnarray}
  F^{\rm def}_{\mu'\nu'}&=&h^{\rm def}_{\mu'\nu'}
+\sum_{i'\in{\rm frag}}\left[2\left(\mu'\nu'|i'i'\right)
  - \left(\mu'i'|i'\nu'\right)\right]\label{eq:defect_F} 
\end{eqnarray}

For the HF energy, the equivalence between the standard HF-embedded-fragment
model \cite{masur2016,christlmaier21} and the model of an aperiodic fragment no
longer holds even if the aperiodic fragment is frozen.  In the former model,
the fragment HF energy {\it was chosen} to coincide with the periodic HF energy per cell, while the model developed in the current work specifies the fragment HF energy as the actual mean-field energy of the physical fragment subjected to embedding. Yet, as is shown in the Supplemental Material, section S1, \cite{suppl} the fragment HF energy can formally be written using the standard HF energy expression using the fragment one-electron Hamiltonian $h^{\rm def}$ and fragment Fock matrix $F^{\rm def}$, but with a redefined effective ``nuclear energy'' $E^{\rm def}_{\rm nuc}$:
\begin{eqnarray}
E^{\rm def}_{\rm HF}&=&{1 \over 2} \left(\sum_{i'\in{\rm
                           frag}}\left[2h^{\rm def}_{i'i'}+2F^{\rm
    def}_{i'i'}\right]\right)
+E^{\rm def}_{\rm nuc},\label{eq:defect_E}
  \end{eqnarray}                         
This $E^{\rm def}_{\rm nuc}$ apart from the repulsion between the fragment nuclei includes the interaction between the latter and the embedding field from the environment:
\begin{eqnarray}
  E^{\rm def}_{\rm nuc}&=&{1 \over 2}\sum_{K'\in{\rm frag}}\sum_{L'\in{\rm frag}}\phantom{}^{'}{Z_{K'}Z_{L'} \over |{\bf R}_{K'}-{\bf
      R}_{L'}|}\nonumber\\
  &&+\sum_{K'\in{\rm frag}}\left[ Z_{K'}\cdot V\left({\bf
                          R}_{K'}\right) -\sum_{K\in{\rm frag}}\phantom{}^{'}{Z_{K}Z_{K'} \over |{\bf R}_{K}-{\bf R}_{K'}|}\right]\nonumber\\
  &&-2\sum_{i\in{\rm frag}}\left< i\left|-\sum_{K'\in{\rm frag}}{Z_{K'} \over |{\bf r}-{\bf  R}_{K'}|}\right|i\right>.\label{eq:E_nuc_defect}
\end{eqnarray}
Here $V\left({\bf R_{K'}}\right)$ is the electrostatic potential of the non-defective periodic system at the locations of the fragment nuclei, and the primed summations exclude the terms with the identical indices. 

Like in eq. (\ref{eq:defect_h}), the summations over $K$ and $K'$ in
(\ref{eq:E_nuc_defect}) are performed over the fragment nuclei before and
after formation of the defect,
respectively. However, it is neither necessary to include the unaltered fragment
nuclei (i.e. when $K$=$K'$) in (\ref{eq:defect_h}) nor in
(\ref{eq:E_nuc_defect}). As demonstrated in Supplemental Material, section S3.A, \cite{suppl} inclusion of
such atoms in the fragment is irrelevant for energy differences, as they add only a constant shift to the defect's HF
energy.  The essential property of the aperiodic defect approach 
is the asymptotic intensivity of energy
differences, as is shown in Supplemental Material (section S3) \cite{suppl}. This makes the physically relevant quantities in our method well-defined and convergent to the TDL.

Having obtained the fragment's HF solution, one can perform a standard
post-HF treatment, single- or multireference, by passing the fragment Hamiltonian matrices $h^{\rm def}_{r's'}$ and $(r's'|t'u')$ in the basis of the fragment HF orbitals $r'$, $s'$, ... (occupied and virtual), as well as $E^{\rm def}_{\rm nuc}$ to a molecular quantum-chemistry program. For this purpose
we employ the FCIDUMP-file interface \cite{Knowles89}, which can be read by a number of molecular codes. In this
work, the post-HF calculations were carried out using Molpro
\cite{molpro}.

Further details on the formalism and the implementation of the aperiodic defect
approach are given in Supplemental Material, sections S1 and S2. The workflow of a standard aperiodic-defect calculation is shown in Fig. S1 \cite{suppl}.

\section{Calculations and discussion}

We begin by pointing out the major differences between
aperiodic-defect model and the standard supercell
approach, in particular the method of Ref. \onlinecite{christlmaier21}, where the fragment is embedded in the mean field of a periodic defect-free calculation. Apart from the obvious efficiency gains at the mean-field part of the
calculation, aperiodic defect offers also conceptual advantages. The
unphysical periodization of a defect in the standard model influences the final result in multiple ways. First, the HF energy includes the interactions between the defect images. If the defect is charged, this causes artifacts due to the presence of the unphysical background, as mentioned above. Even if the defect is neutral but polar, the slowly decaying dipole-dipole interactions considerably slow down the convergence to the TDL with respect to the supercell size. Second, the post-HF fragment treatment involves a second parameter -- the fragment size -- with respect to which the result has to be converged. Furthermore, since the in-fragment treatment is coupled to the supercell calculation via the reference state (i.e. the periodic HF determinant), the embedding field, the choice of the fragment orbitals, etc., the convergence with the supercell size in the fragment-based post-HF calculations may become even slower.

In contrast to that in the aperiodic defect model, there is only one parameter  -- the size of the fragment -- that governs the finite size error. And its convergence to the TDL is expected to be much faster, as the embedding field and the reference function do not contain the accumulated electrostatics or static-correlation artifacts. The source of the finite size error in this model is merely the lack of the {\it induced} electrostatic and van der Waals dispersive interactions between the fragment and the environment. Both these effects are usually of a shorter range and a smaller magnitude than for example dipole-dipole interactions between the defect images.

As a test system, we have taken graphane, which is 2D-periodic and nonconducting, with the defect being a substitution of a hydrogen atom with fluorine (henceforth referred to as fluorographane). The technical details of the calculations are outlined in the Supplemental Material, section S4 \cite{suppl}. The aperiodic defect model of fluorographane with a possible choice of fragment is shown in Figure S2 of the Supplemental Material \cite{suppl}.
Fluorographane is a convenient testbed for our method, as it has a substantial amount of dynamic correlation from valence-electron-rich fluorine, and, at the same time, one can tune the amount of static correlation by a gradual dissociation of the fluorine atom. Systems with both dynamic (weak) and static (strong) correlation, a situation common for defects, are generally difficult for electronic structure theory. DFT may fail 
for such systems even qualitatively. Quantum chemical single-reference methods, while adept at treating dynamic correlation with systematic improvability also fail for static correlation (with rare exceptions). Multireference methods capture static correlation, but generally struggle for large systems to affordably treat the dynamic correlation. Finally, we note that in addition to the correlation effects in fluorographane, the ionic dissociation state appearing in the Hartree-Fock description generates a large dipole moment on the defect, making our system also a prototype of a strongly polar defect.

\begin{figure}[H]
    \centering
    \includegraphics[width=0.45\textwidth]{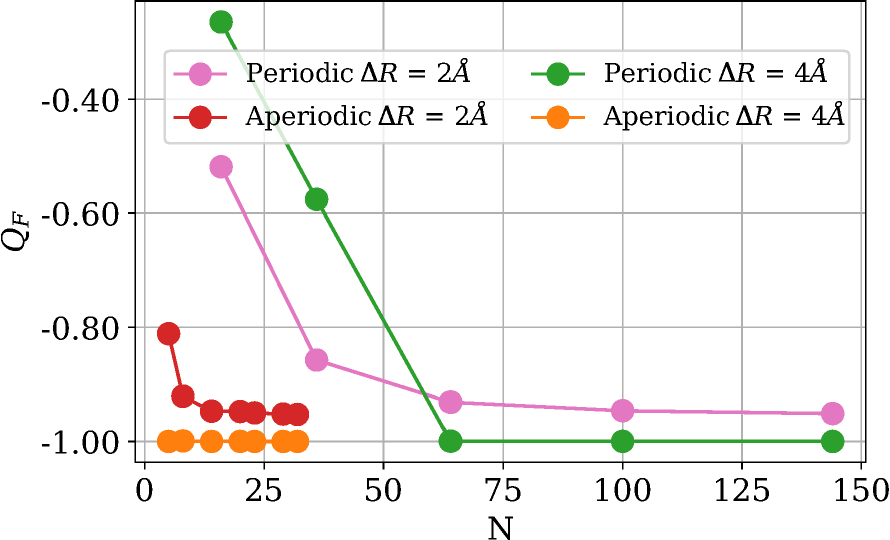}
    \caption{The HF Mulliken charge of the fluorine atom $Q_F$ in Fluorographane at 2 {\AA} and 4 {\AA} of the C-F bond elongation as a function of the number of atoms N in the fragment for the aperiodic defect model or in the supercell for the periodic defect model}
    \label{fig:mulliken}
\end{figure}

We start off with the analysis of the accumulation of static correlation in the periodic model due to the multiple C-F bond-breaking in the periodic model. Fig. \ref{fig:mulliken} shows how the Mulliken charge on the F atoms converges to its TDL value. For single C-F bond dissociation HF gives an excited fully ionic state, which is intrinsically single-reference. Therefore at the TDL, the Mulliken charge on F at dissociation is expected to be very close to -1. This is very well reproduced by the aperiodic model already with very small fragments. For the periodic defect model, however, this is achieved only at quite large supercells. For the 2{\AA} elongation, for example, the converged value is not recovered until the 100-atomic supercell.

With smaller supercells the lower charge on fluorine atoms reveals a mixed ionic/neutral state dissociation -- the textbook HF marker for static correlation, revealing the collective delocalizating effect of multiple bond-breaking in the periodic defect model. This effect also led to the breakdown of the MP2 potential energy curve in Ref. \onlinecite{christlmaier21} indicating an essential multireference character. In contrast, in the aperiodic defect model, the MP2 curve follows the HF one without deterioration, see Fig S3 in the Supplemental Material \cite{suppl}, confirming a well behaved single-reference state. Then the truly multireference ground state corresponding to the neutral dissociation is reproduced in the aperiodic defect model by the proper multireference treatment (vide infra) or the Brueckner distinguishable cluster doubles (BDCD)  \cite{Kats2013,Kats2014} method, see section S7 of the Supplemental Material \cite{suppl}.

\begin{figure}[H]
    \centering
    \includegraphics[width=0.45\textwidth]{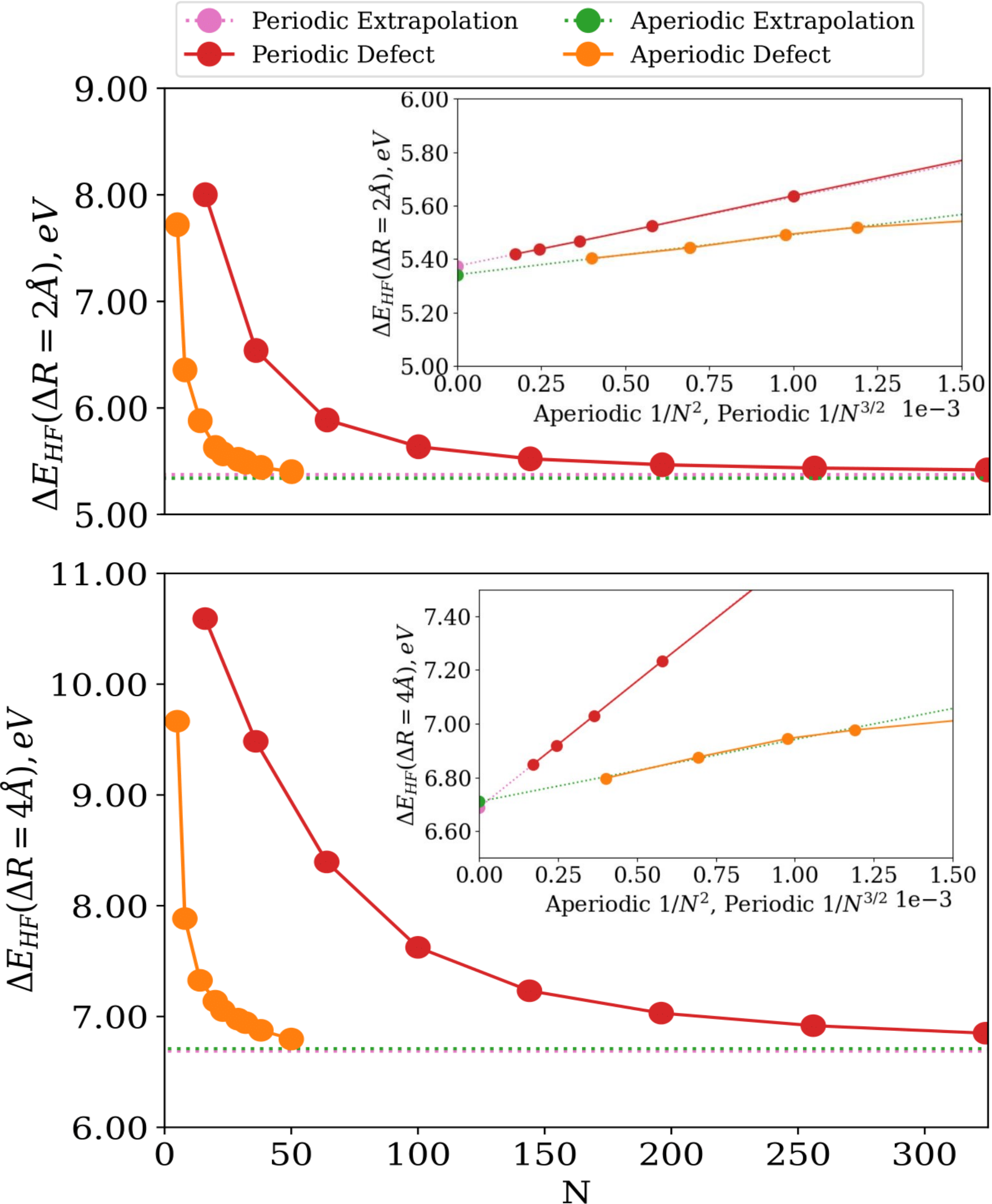}
    \caption{The HF dissociation energy for the fluorine atom in fluorographane corresponding to bond elongations of 2 {\AA} and 4 {\AA} as functions of N, where N is number of atoms in the fragment for the aperiodic defect model and number of atoms in the supercell for the periodic model. In the insets, the dissociation energies are given as functions of $1/N^{3/2}$ for the periodic model and $1/N^2$ for the aperiodic model with a linear (in these coordinates) extrapolation to TDL, see section S6 of the Supplemental Material \cite{suppl} for analysis of the asymptotic behavior. The horizontal lines indicate the extrapolated values.}
    \label{fig:HF_TDL}
\end{figure}

Next we focus on the convergence of the HF dissociation energies to the TDL. It is evident from Fig. \ref{fig:HF_TDL} how much faster the aperiodic model converges. For example, the aperiodic model with a 50-atom fragment is closer to the TDL than a 325-atom supercell calculation, as the dipole-dipole interactions in the latter are of a very long range and persist with very large supercells. We note that, in a 3D-periodic system a supercell of such a size would incorporate thousands of atoms, making calculations hardly feasible. Importantly, the TDL-extrapolated dissociation energies of the aperiodic model coincide (up to an extrapolation error) with those of the fully periodic model, illustrating the asymptotic size-intensivity for energy differences of the former, as discussed above.

\begin{figure}[H]
    \centering
    \includegraphics[width=0.45\textwidth]{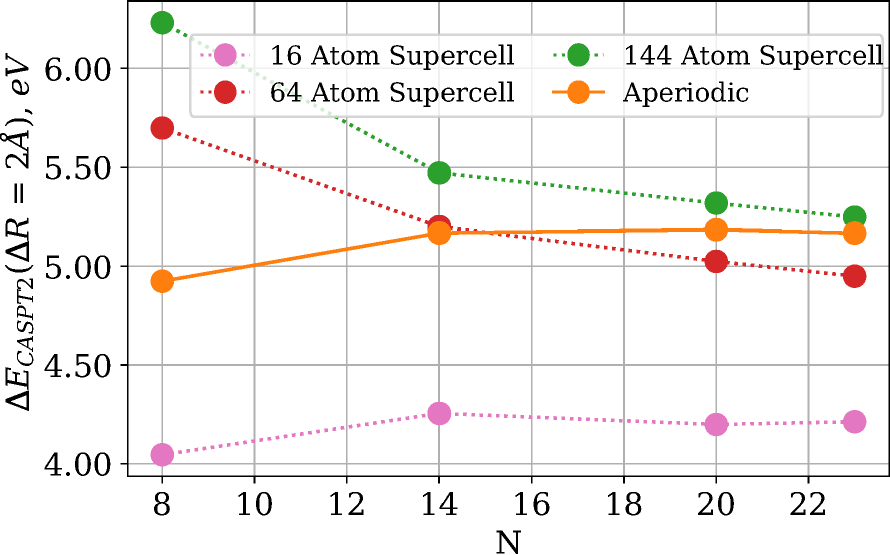}
    \caption{The CASPT2(6,7) dissociation energy for the fluorine atom in fluorographane corresponding to the bond elongation of 2{\AA} as a function of the number of atoms in the fragment. }
    \label{fig:CASPT2}
\end{figure}

Finally we investigate the actual ground state of the fluorine dissociation. It corresponds to the neutral dissociation and is truly multireference. 
Fig. \ref{fig:CASPT2} presents the CASPT2 dissociation energies calculated with the periodic and aperiodic models. It is evident that in the aperiodic model the dissociation energy converges with the fragment size very quickly. As the bond elongation does not create a dipole moment, no extrapolation is actually needed here.

As mentioned above, the periodic embedded fragment model requires to converge two parameters to TDL: the supercell size and the fragment size. Despite the neutral dissociation inside the fragment described by CASPT2, the embedding field there is still contaminated by the defect images treated at the HF level, with the involved strong polarization, delocalized static correlation, etc. This makes the convergence in the periodic model with the supercell size and within a given supercell with the fragment size very slow.  It is clear that with the expansion of the supercell the periodic results are getting closer to those of the aperiodic model, but even the largest supercell tested, consisting of 144 atoms, is not enough for convergence.

The slow convergence with the supercell and fragment sizes is not the periodic model's only problem. If the supercell is not large enough, the collective delocalized static correlation causes convergence problems in the periodic SCF and, especially so, in construction of localized orbitals needed for the fragment definition. These two steps become progressively unstable with the bond elongation, effectively prohibiting the use of the periodic fragment approach beyond 2 {\AA}, at least for small and medium sized supercells. In contrast, with the aperiodic defect method, the periodic HF and localization are
unproblematic and correspond to the well-behaved, closed-shell, non-defective
system regardless of the a-posteriori bond elongation
distance.

\section{Conclusions}

In this Letter, we have introduced a quantum embedding scheme for an ``aperiodic'' fragment embedded in the ``defectless'' periodic HF solution, eliminating the need for periodic repetition of the defect and associated expensive supercells, compensating background charges, etc. in electronic structure calculations of solid-state defects. An aperiodic fragment containing the defect is
 placed in a frozen environment of the periodic Hartree Fock solution calculated using pristine, minimal unit cells, leading to significant
 computational savings and reduced approximations. The only required
 assumption is that the fragment is large enough to capture the electronic
 structure of the defect and the response from its surroundings, such that the TDL limit values for relevant energy differences can be obtained either directly or via extrapolation.

The stark contrast to the periodic model shows the importance of
using aperiodic approach for polarized, charged or/and strongly correlated defects, as otherwise a comparable accuracy may only be reached with very large supercells in much more computationally demanding, if feasible at all, calculations. Demonstration of performance of the aperiodic defect model for realistic treatment of charged defects will be shown in a separate publication.

  While we formulate this approach in the context of
 periodic Hartree Fock embedding, it can be readily extended to
 other mean-field approaches. HF embedding with DFT orbitals is
 possible right away without any additional implementation efforts. A proper DFT embedding would require calculation of the integrals
 for the exchange-correlation potential, but otherwise very little effort (Fock exchange would merely be supplemented or replaced by exchange-correlation). As in other DFT embedding models, it remains to be seen if a double-counting of non-additive components would represent an issue. 
Another possible extension of the scheme is to increase the level of embedding, for example by incorporation of the explicit Coulomb response from the
environment \cite{WachterLehn2022}, which may speed up the
convergence with fragment size even further. Transcorrelation approaches \cite{Cohen2019,Christlmaier2023} could also assist in reducing the basis set error and improving the overall accuracy.


\section{Acknowledgements}

R.H.L. acknowledges support by the NSF Graduate Research Fellowship under Grant No. DGE-2034835, as well as an abroad, collaborative effort with D.U. made possible by the NSF IRES program under Grant No. OISE-1658652. A.N.A. acknowledges the support of the NSF Center for Chemical Innovation Phase I (Grant No. CHE-2221453).


\typeout{}
\bibliography{citations}

\begin{thebibliography}{90}%
\makeatletter
\providecommand \@ifxundefined [1]{%
 \@ifx{#1\undefined}
}%
\providecommand \@ifnum [1]{%
 \ifnum #1\expandafter \@firstoftwo
 \else \expandafter \@secondoftwo
 \fi
}%
\providecommand \@ifx [1]{%
 \ifx #1\expandafter \@firstoftwo
 \else \expandafter \@secondoftwo
 \fi
}%
\providecommand \natexlab [1]{#1}%
\providecommand \enquote  [1]{``#1''}%
\providecommand \bibnamefont  [1]{#1}%
\providecommand \bibfnamefont [1]{#1}%
\providecommand \citenamefont [1]{#1}%
\providecommand \href@noop [0]{\@secondoftwo}%
\providecommand \href [0]{\begingroup \@sanitize@url \@href}%
\providecommand \@href[1]{\@@startlink{#1}\@@href}%
\providecommand \@@href[1]{\endgroup#1\@@endlink}%
\providecommand \@sanitize@url [0]{\catcode `\\12\catcode `\$12\catcode `\&12\catcode `\#12\catcode `\^12\catcode `\_12\catcode `\%12\relax}%
\providecommand \@@startlink[1]{}%
\providecommand \@@endlink[0]{}%
\providecommand \url  [0]{\begingroup\@sanitize@url \@url }%
\providecommand \@url [1]{\endgroup\@href {#1}{\urlprefix }}%
\providecommand \urlprefix  [0]{URL }%
\providecommand \Eprint [0]{\href }%
\providecommand \doibase [0]{https://doi.org/}%
\providecommand \selectlanguage [0]{\@gobble}%
\providecommand \bibinfo  [0]{\@secondoftwo}%
\providecommand \bibfield  [0]{\@secondoftwo}%
\providecommand \translation [1]{[#1]}%
\providecommand \BibitemOpen [0]{}%
\providecommand \bibitemStop [0]{}%
\providecommand \bibitemNoStop [0]{.\EOS\space}%
\providecommand \EOS [0]{\spacefactor3000\relax}%
\providecommand \BibitemShut  [1]{\csname bibitem#1\endcsname}%
\let\auto@bib@innerbib\@empty
\bibitem [{\citenamefont {Pisani}\ \emph {et~al.}(2008)\citenamefont {Pisani}, \citenamefont {Maschio}, \citenamefont {Casassa}, \citenamefont {Halo}, \citenamefont {Sch\"utz},\ and\ \citenamefont {Usvyat}}]{pisani2008}%
  \BibitemOpen
  \bibfield  {author} {\bibinfo {author} {\bibfnamefont {C.}~\bibnamefont {Pisani}}, \bibinfo {author} {\bibfnamefont {L.}~\bibnamefont {Maschio}}, \bibinfo {author} {\bibfnamefont {S.}~\bibnamefont {Casassa}}, \bibinfo {author} {\bibfnamefont {M.}~\bibnamefont {Halo}}, \bibinfo {author} {\bibfnamefont {M.}~\bibnamefont {Sch\"utz}},\ and\ \bibinfo {author} {\bibfnamefont {D.}~\bibnamefont {Usvyat}},\ }\bibfield  {title} {\bibinfo {title} {Periodic local mp2 method for the study of electronic correlation in crystals: Theory and preliminary applications},\ }\href@noop {} {\bibfield  {journal} {\bibinfo  {journal} {J. Comput. Chem.}\ }\textbf {\bibinfo {volume} {29}},\ \bibinfo {pages} {2113} (\bibinfo {year} {2008})}\BibitemShut {NoStop}%
\bibitem [{\citenamefont {Marsman}\ \emph {et~al.}(2009)\citenamefont {Marsman}, \citenamefont {Grüneis}, \citenamefont {Paier},\ and\ \citenamefont {Kresse}}]{Marsman2009}%
  \BibitemOpen
  \bibfield  {author} {\bibinfo {author} {\bibfnamefont {M.}~\bibnamefont {Marsman}}, \bibinfo {author} {\bibfnamefont {A.}~\bibnamefont {Grüneis}}, \bibinfo {author} {\bibfnamefont {J.}~\bibnamefont {Paier}},\ and\ \bibinfo {author} {\bibfnamefont {G.}~\bibnamefont {Kresse}},\ }\bibfield  {title} {\bibinfo {title} {Second-order møller-plesset perturbation theory applied to extended systems. i. within the projector-augmented-wave formalism using a plane wave basis set},\ }\href@noop {} {\bibfield  {journal} {\bibinfo  {journal} {J. Chem. Phys.}\ }\textbf {\bibinfo {volume} {130}},\ \bibinfo {pages} {184103} (\bibinfo {year} {2009})}\BibitemShut {NoStop}%
\bibitem [{\citenamefont {Booth}\ \emph {et~al.}(2013)\citenamefont {Booth}, \citenamefont {Grüneis}, \citenamefont {Kresse},\ and\ \citenamefont {Alavi}}]{Booth2013_}%
  \BibitemOpen
  \bibfield  {author} {\bibinfo {author} {\bibfnamefont {G.~H.}\ \bibnamefont {Booth}}, \bibinfo {author} {\bibfnamefont {A.}~\bibnamefont {Grüneis}}, \bibinfo {author} {\bibfnamefont {G.}~\bibnamefont {Kresse}},\ and\ \bibinfo {author} {\bibfnamefont {A.}~\bibnamefont {Alavi}},\ }\bibfield  {title} {\bibinfo {title} {Towards an exact description of electronic wavefunctions in real solids},\ }\href@noop {} {\bibfield  {journal} {\bibinfo  {journal} {Nature}\ }\textbf {\bibinfo {volume} {493}},\ \bibinfo {pages} {365} (\bibinfo {year} {2013})}\BibitemShut {NoStop}%
\bibitem [{\citenamefont {Usvyat}\ \emph {et~al.}(2015)\citenamefont {Usvyat}, \citenamefont {Maschio},\ and\ \citenamefont {Sch\"{u}tz}}]{usvyat2015}%
  \BibitemOpen
  \bibfield  {author} {\bibinfo {author} {\bibfnamefont {D.}~\bibnamefont {Usvyat}}, \bibinfo {author} {\bibfnamefont {L.}~\bibnamefont {Maschio}},\ and\ \bibinfo {author} {\bibfnamefont {M.}~\bibnamefont {Sch\"{u}tz}},\ }\bibfield  {title} {\bibinfo {title} {{Periodic Local MP2 Method Employing Orbital Specific Virtuals}},\ }\href@noop {} {\bibfield  {journal} {\bibinfo  {journal} {J. Chem. Phys}\ }\textbf {\bibinfo {volume} {143}},\ \bibinfo {pages} {102805} (\bibinfo {year} {2015})}\BibitemShut {NoStop}%
\bibitem [{\citenamefont {Ben}\ \emph {et~al.}(2013)\citenamefont {Ben}, \citenamefont {Hutter},\ and\ \citenamefont {Vandevondele}}]{delben2}%
  \BibitemOpen
  \bibfield  {author} {\bibinfo {author} {\bibfnamefont {M.~D.}\ \bibnamefont {Ben}}, \bibinfo {author} {\bibfnamefont {J.}~\bibnamefont {Hutter}},\ and\ \bibinfo {author} {\bibfnamefont {J.}~\bibnamefont {Vandevondele}},\ }\bibfield  {title} {\bibinfo {title} {Electron correlation in the condensed phase from a resolution of identity approach based on the gaussian and plane waves scheme},\ }\href@noop {} {\bibfield  {journal} {\bibinfo  {journal} {J. Chem. Theory Comput.}\ }\textbf {\bibinfo {volume} {9}},\ \bibinfo {pages} {2654} (\bibinfo {year} {2013})}\BibitemShut {NoStop}%
\bibitem [{\citenamefont {Booth}\ \emph {et~al.}(2016)\citenamefont {Booth}, \citenamefont {Tsatsoulis}, \citenamefont {Chan},\ and\ \citenamefont {Grüneis}}]{Booth2016}%
  \BibitemOpen
  \bibfield  {author} {\bibinfo {author} {\bibfnamefont {G.~H.}\ \bibnamefont {Booth}}, \bibinfo {author} {\bibfnamefont {T.}~\bibnamefont {Tsatsoulis}}, \bibinfo {author} {\bibfnamefont {G.~K.~L.}\ \bibnamefont {Chan}},\ and\ \bibinfo {author} {\bibfnamefont {A.}~\bibnamefont {Grüneis}},\ }\bibfield  {title} {\bibinfo {title} {From plane waves to local gaussians for the simulation of correlated periodic systems},\ }\href@noop {} {\bibfield  {journal} {\bibinfo  {journal} {J. Chem. Phys.}\ }\textbf {\bibinfo {volume} {145}},\ \bibinfo {pages} {084111} (\bibinfo {year} {2016})}\BibitemShut {NoStop}%
\bibitem [{\citenamefont {McClain}\ \emph {et~al.}(2017)\citenamefont {McClain}, \citenamefont {Sun}, \citenamefont {Chan},\ and\ \citenamefont {Berkelbach}}]{McClain2017}%
  \BibitemOpen
  \bibfield  {author} {\bibinfo {author} {\bibfnamefont {J.}~\bibnamefont {McClain}}, \bibinfo {author} {\bibfnamefont {Q.}~\bibnamefont {Sun}}, \bibinfo {author} {\bibfnamefont {G.~K.-L.}\ \bibnamefont {Chan}},\ and\ \bibinfo {author} {\bibfnamefont {T.~C.}\ \bibnamefont {Berkelbach}},\ }\bibfield  {title} {\bibinfo {title} {Gaussian-based coupled-cluster theory for the ground-state and band structure of solids},\ }\href@noop {} {\bibfield  {journal} {\bibinfo  {journal} {J. Chem. Theory Comput.}\ }\textbf {\bibinfo {volume} {13}},\ \bibinfo {pages} {1209} (\bibinfo {year} {2017})}\BibitemShut {NoStop}%
\bibitem [{\citenamefont {Gruber}\ \emph {et~al.}(2018)\citenamefont {Gruber}, \citenamefont {Liao}, \citenamefont {Tsatsoulis}, \citenamefont {Hummel},\ and\ \citenamefont {Gr{\"u}neis}}]{grueneis2018}%
  \BibitemOpen
  \bibfield  {author} {\bibinfo {author} {\bibfnamefont {T.}~\bibnamefont {Gruber}}, \bibinfo {author} {\bibfnamefont {K.}~\bibnamefont {Liao}}, \bibinfo {author} {\bibfnamefont {T.}~\bibnamefont {Tsatsoulis}}, \bibinfo {author} {\bibfnamefont {F.}~\bibnamefont {Hummel}},\ and\ \bibinfo {author} {\bibfnamefont {A.}~\bibnamefont {Gr{\"u}neis}},\ }\bibfield  {title} {\bibinfo {title} {Applying the coupled-cluster ansatz to solids and surfaces in the thermodynamic limit},\ }\href@noop {} {\bibfield  {journal} {\bibinfo  {journal} {Phys. Rev. X}\ }\textbf {\bibinfo {volume} {8}},\ \bibinfo {pages} {021043} (\bibinfo {year} {2018})}\BibitemShut {NoStop}%
\bibitem [{\citenamefont {Liao}\ \emph {et~al.}(2021)\citenamefont {Liao}, \citenamefont {Schraivogel}, \citenamefont {Luo}, \citenamefont {Kats},\ and\ \citenamefont {Alavi}}]{Liao2021}%
  \BibitemOpen
  \bibfield  {author} {\bibinfo {author} {\bibfnamefont {K.}~\bibnamefont {Liao}}, \bibinfo {author} {\bibfnamefont {T.}~\bibnamefont {Schraivogel}}, \bibinfo {author} {\bibfnamefont {H.}~\bibnamefont {Luo}}, \bibinfo {author} {\bibfnamefont {D.}~\bibnamefont {Kats}},\ and\ \bibinfo {author} {\bibfnamefont {A.}~\bibnamefont {Alavi}},\ }\bibfield  {title} {\bibinfo {title} {Towards efficient and accurate ab initio solutions to periodic systems via transcorrelation and coupled cluster theory},\ }\href@noop {} {\bibfield  {journal} {\bibinfo  {journal} {Phys. Rev. Res.}\ }\textbf {\bibinfo {volume} {3}},\ \bibinfo {pages} {033072} (\bibinfo {year} {2021})}\BibitemShut {NoStop}%
\bibitem [{\citenamefont {Neufeld}\ and\ \citenamefont {Berkelbach}(2023)}]{Neufeld2023}%
  \BibitemOpen
  \bibfield  {author} {\bibinfo {author} {\bibfnamefont {V.~A.}\ \bibnamefont {Neufeld}}\ and\ \bibinfo {author} {\bibfnamefont {T.~C.}\ \bibnamefont {Berkelbach}},\ }\bibfield  {title} {\bibinfo {title} {Highly accurate electronic structure of metallic solids from coupled-cluster theory with nonperturbative triple excitations},\ }\href@noop {} {\bibfield  {journal} {\bibinfo  {journal} {Phys. Rev. Lett.}\ }\textbf {\bibinfo {volume} {131}},\ \bibinfo {pages} {186402} (\bibinfo {year} {2023})}\BibitemShut {NoStop}%
\bibitem [{\citenamefont {Stoll}(1992)}]{stoll1992}%
  \BibitemOpen
  \bibfield  {author} {\bibinfo {author} {\bibfnamefont {H.}~\bibnamefont {Stoll}},\ }\bibfield  {title} {\bibinfo {title} {Correlation energy of diamond},\ }\href@noop {} {\bibfield  {journal} {\bibinfo  {journal} {Phys. Rev. B}\ }\textbf {\bibinfo {volume} {46}},\ \bibinfo {pages} {6700} (\bibinfo {year} {1992})}\BibitemShut {NoStop}%
\bibitem [{\citenamefont {Paulus}(2006)}]{paulus2006}%
  \BibitemOpen
  \bibfield  {author} {\bibinfo {author} {\bibfnamefont {B.}~\bibnamefont {Paulus}},\ }\bibfield  {title} {\bibinfo {title} {The method of increments --- a wavefunction-based ab initio correlation method for solids},\ }\href@noop {} {\bibfield  {journal} {\bibinfo  {journal} {Phys. Rep.}\ }\textbf {\bibinfo {volume} {428}},\ \bibinfo {pages} {1} (\bibinfo {year} {2006})}\BibitemShut {NoStop}%
\bibitem [{\citenamefont {Tuma}\ and\ \citenamefont {Sauer}(2006)}]{Tuma2006}%
  \BibitemOpen
  \bibfield  {author} {\bibinfo {author} {\bibfnamefont {C.}~\bibnamefont {Tuma}}\ and\ \bibinfo {author} {\bibfnamefont {J.}~\bibnamefont {Sauer}},\ }\bibfield  {title} {\bibinfo {title} {Treating dispersion effects in extended systems by hybrid mp2:dft calculations—protonation of isobutene in zeolite ferrierite},\ }\href@noop {} {\bibfield  {journal} {\bibinfo  {journal} {Phys. Chem. Chem. Phys.}\ }\textbf {\bibinfo {volume} {8}},\ \bibinfo {pages} {3955} (\bibinfo {year} {2006})}\BibitemShut {NoStop}%
\bibitem [{\citenamefont {Nolan}\ \emph {et~al.}(2009)\citenamefont {Nolan}, \citenamefont {Gillan}, \citenamefont {Alf{\`e}}, \citenamefont {Allan},\ and\ \citenamefont {Manby}}]{nolan2009}%
  \BibitemOpen
  \bibfield  {author} {\bibinfo {author} {\bibfnamefont {S.~J.}\ \bibnamefont {Nolan}}, \bibinfo {author} {\bibfnamefont {M.~J.}\ \bibnamefont {Gillan}}, \bibinfo {author} {\bibfnamefont {D.}~\bibnamefont {Alf{\`e}}}, \bibinfo {author} {\bibfnamefont {N.~L.}\ \bibnamefont {Allan}},\ and\ \bibinfo {author} {\bibfnamefont {F.~R.}\ \bibnamefont {Manby}},\ }\bibfield  {title} {\bibinfo {title} {Calculation of properties of crystalline lithium hydride using correlated wave function theory},\ }\href@noop {} {\bibfield  {journal} {\bibinfo  {journal} {Phys. Rev. B}\ }\textbf {\bibinfo {volume} {80}},\ \bibinfo {pages} {165109} (\bibinfo {year} {2009})}\BibitemShut {NoStop}%
\bibitem [{\citenamefont {Wen}\ and\ \citenamefont {Beran}(2011)}]{Wen2011}%
  \BibitemOpen
  \bibfield  {author} {\bibinfo {author} {\bibfnamefont {S.}~\bibnamefont {Wen}}\ and\ \bibinfo {author} {\bibfnamefont {G.~J.}\ \bibnamefont {Beran}},\ }\bibfield  {title} {\bibinfo {title} {Accurate molecular crystal lattice energies from a fragment qm/mm approach with on-the-fly ab initio force field parametrization},\ }\href@noop {} {\bibfield  {journal} {\bibinfo  {journal} {J. Chem. Theory Comput.}\ }\textbf {\bibinfo {volume} {7}},\ \bibinfo {pages} {3733} (\bibinfo {year} {2011})}\BibitemShut {NoStop}%
\bibitem [{\citenamefont {Yang}\ \emph {et~al.}(2014)\citenamefont {Yang}, \citenamefont {Hu}, \citenamefont {Usvyat}, \citenamefont {Matthews}, \citenamefont {Sch{\"u}tz},\ and\ \citenamefont {Chan}}]{Chan_science}%
  \BibitemOpen
  \bibfield  {author} {\bibinfo {author} {\bibfnamefont {J.}~\bibnamefont {Yang}}, \bibinfo {author} {\bibfnamefont {W.}~\bibnamefont {Hu}}, \bibinfo {author} {\bibfnamefont {D.}~\bibnamefont {Usvyat}}, \bibinfo {author} {\bibfnamefont {D.}~\bibnamefont {Matthews}}, \bibinfo {author} {\bibfnamefont {M.}~\bibnamefont {Sch{\"u}tz}},\ and\ \bibinfo {author} {\bibfnamefont {G.~K.-L.}\ \bibnamefont {Chan}},\ }\bibfield  {title} {\bibinfo {title} {Ab initio determination of the crystalline benzene lattice energy to sub-kilojoule/mole accuracy},\ }\href@noop {} {\bibfield  {journal} {\bibinfo  {journal} {Science}\ }\textbf {\bibinfo {volume} {345}},\ \bibinfo {pages} {6197} (\bibinfo {year} {2014})}\BibitemShut {NoStop}%
\bibitem [{\citenamefont {Sch\"utz}\ \emph {et~al.}(2017)\citenamefont {Sch\"utz}, \citenamefont {Maschio}, \citenamefont {Karttunen},\ and\ \citenamefont {Usvyat}}]{schutz2017}%
  \BibitemOpen
  \bibfield  {author} {\bibinfo {author} {\bibfnamefont {M.}~\bibnamefont {Sch\"utz}}, \bibinfo {author} {\bibfnamefont {L.}~\bibnamefont {Maschio}}, \bibinfo {author} {\bibfnamefont {A.~J.}\ \bibnamefont {Karttunen}},\ and\ \bibinfo {author} {\bibfnamefont {D.}~\bibnamefont {Usvyat}},\ }\bibfield  {title} {\bibinfo {title} {Exfoliation energy of black phosphorus revisited: A coupled cluster benchmark},\ }\href@noop {} {\bibfield  {journal} {\bibinfo  {journal} {J. Phys. Chem. Lett.}\ }\textbf {\bibinfo {volume} {8}},\ \bibinfo {pages} {1290} (\bibinfo {year} {2017})}\BibitemShut {NoStop}%
\bibitem [{\citenamefont {Alessio}\ \emph {et~al.}(2018)\citenamefont {Alessio}, \citenamefont {Bischoff},\ and\ \citenamefont {Sauer}}]{alessio18}%
  \BibitemOpen
  \bibfield  {author} {\bibinfo {author} {\bibfnamefont {M.}~\bibnamefont {Alessio}}, \bibinfo {author} {\bibfnamefont {F.~A.}\ \bibnamefont {Bischoff}},\ and\ \bibinfo {author} {\bibfnamefont {J.}~\bibnamefont {Sauer}},\ }\bibfield  {title} {\bibinfo {title} {Chemically accurate adsorption energies for methane and ethane monolayers on the mgo (001) surface},\ }\href@noop {} {\bibfield  {journal} {\bibinfo  {journal} {Phys. Chem. Chem. Phys.}\ }\textbf {\bibinfo {volume} {20}},\ \bibinfo {pages} {9760} (\bibinfo {year} {2018})}\BibitemShut {NoStop}%
\bibitem [{\citenamefont {Govind}\ \emph {et~al.}(1998)\citenamefont {Govind}, \citenamefont {Wang}, \citenamefont {Silva},\ and\ \citenamefont {Carter}}]{Govind1998}%
  \BibitemOpen
  \bibfield  {author} {\bibinfo {author} {\bibfnamefont {N.}~\bibnamefont {Govind}}, \bibinfo {author} {\bibfnamefont {Y.~A.}\ \bibnamefont {Wang}}, \bibinfo {author} {\bibfnamefont {A.~J.~D.}\ \bibnamefont {Silva}},\ and\ \bibinfo {author} {\bibfnamefont {E.~A.}\ \bibnamefont {Carter}},\ }\bibfield  {title} {\bibinfo {title} {Accurate ab initio energetics of extended systems via explicit correlation embedded in a density functional environment},\ }\href@noop {} {\bibfield  {journal} {\bibinfo  {journal} {Chem. Phys. Lett.}\ }\textbf {\bibinfo {volume} {295}},\ \bibinfo {pages} {129} (\bibinfo {year} {1998})}\BibitemShut {NoStop}%
\bibitem [{\citenamefont {Klüner}\ \emph {et~al.}(2002)\citenamefont {Klüner}, \citenamefont {Govind}, \citenamefont {Wang},\ and\ \citenamefont {Carter}}]{Kluner}%
  \BibitemOpen
  \bibfield  {author} {\bibinfo {author} {\bibfnamefont {T.}~\bibnamefont {Klüner}}, \bibinfo {author} {\bibfnamefont {N.}~\bibnamefont {Govind}}, \bibinfo {author} {\bibfnamefont {Y.~A.}\ \bibnamefont {Wang}},\ and\ \bibinfo {author} {\bibfnamefont {E.~A.}\ \bibnamefont {Carter}},\ }\bibfield  {title} {\bibinfo {title} {Periodic density functional embedding theory for complete active space self-consistent field and configuration interaction calculations: Ground and excited states},\ }\href@noop {} {\bibfield  {journal} {\bibinfo  {journal} {J. Chem. Phys.}\ }\textbf {\bibinfo {volume} {116}},\ \bibinfo {pages} {42} (\bibinfo {year} {2002})}\BibitemShut {NoStop}%
\bibitem [{\citenamefont {Jacob}\ \emph {et~al.}(2008)\citenamefont {Jacob}, \citenamefont {Neugebauer},\ and\ \citenamefont {Visscher}}]{Jacob08}%
  \BibitemOpen
  \bibfield  {author} {\bibinfo {author} {\bibfnamefont {C.~R.}\ \bibnamefont {Jacob}}, \bibinfo {author} {\bibfnamefont {J.}~\bibnamefont {Neugebauer}},\ and\ \bibinfo {author} {\bibfnamefont {L.}~\bibnamefont {Visscher}},\ }\bibfield  {title} {\bibinfo {title} {A flexible implementation of frozen-density embedding for use in multilevel simulations},\ }\href@noop {} {\bibfield  {journal} {\bibinfo  {journal} {J. Comp. Chem.}\ }\textbf {\bibinfo {volume} {29}},\ \bibinfo {pages} {1011} (\bibinfo {year} {2008})}\BibitemShut {NoStop}%
\bibitem [{\citenamefont {Bygrave}\ \emph {et~al.}(2012)\citenamefont {Bygrave}, \citenamefont {Allan},\ and\ \citenamefont {Manby}}]{bygrave2012}%
  \BibitemOpen
  \bibfield  {author} {\bibinfo {author} {\bibfnamefont {P.~J.}\ \bibnamefont {Bygrave}}, \bibinfo {author} {\bibfnamefont {N.~L.}\ \bibnamefont {Allan}},\ and\ \bibinfo {author} {\bibfnamefont {F.~R.}\ \bibnamefont {Manby}},\ }\bibfield  {title} {\bibinfo {title} {The embedded many-body expansion for energetics of molecular crystals},\ }\href@noop {} {\bibfield  {journal} {\bibinfo  {journal} {J. Chem. Phys.}\ }\textbf {\bibinfo {volume} {137}},\ \bibinfo {pages} {164102} (\bibinfo {year} {2012})}\BibitemShut {NoStop}%
\bibitem [{\citenamefont {Manby}\ \emph {et~al.}(2012)\citenamefont {Manby}, \citenamefont {Stella}, \citenamefont {Goodpaster},\ and\ \citenamefont {Miller}}]{Manby2012}%
  \BibitemOpen
  \bibfield  {author} {\bibinfo {author} {\bibfnamefont {F.~R.}\ \bibnamefont {Manby}}, \bibinfo {author} {\bibfnamefont {M.}~\bibnamefont {Stella}}, \bibinfo {author} {\bibfnamefont {J.~D.}\ \bibnamefont {Goodpaster}},\ and\ \bibinfo {author} {\bibfnamefont {T.~F.}\ \bibnamefont {Miller}},\ }\bibfield  {title} {\bibinfo {title} {A simple, exact density-functional-theory embedding scheme},\ }\href@noop {} {\bibfield  {journal} {\bibinfo  {journal} {J. Chem. Theory Comput.}\ }\textbf {\bibinfo {volume} {8}},\ \bibinfo {pages} {2564} (\bibinfo {year} {2012})}\BibitemShut {NoStop}%
\bibitem [{\citenamefont {Libisch}\ \emph {et~al.}(2014)\citenamefont {Libisch}, \citenamefont {Huang},\ and\ \citenamefont {Carter}}]{Libisch2014}%
  \BibitemOpen
  \bibfield  {author} {\bibinfo {author} {\bibfnamefont {F.}~\bibnamefont {Libisch}}, \bibinfo {author} {\bibfnamefont {C.}~\bibnamefont {Huang}},\ and\ \bibinfo {author} {\bibfnamefont {E.~A.}\ \bibnamefont {Carter}},\ }\bibfield  {title} {\bibinfo {title} {Embedded correlated wavefunction schemes: Theory and applications},\ }\href@noop {} {\bibfield  {journal} {\bibinfo  {journal} {Acc. Chem. Res.}\ }\textbf {\bibinfo {volume} {47}},\ \bibinfo {pages} {2768} (\bibinfo {year} {2014})}\BibitemShut {NoStop}%
\bibitem [{\citenamefont {Sun}\ and\ \citenamefont {Chan}(2016)}]{Sun2016}%
  \BibitemOpen
  \bibfield  {author} {\bibinfo {author} {\bibfnamefont {Q.}~\bibnamefont {Sun}}\ and\ \bibinfo {author} {\bibfnamefont {G.~K.~L.}\ \bibnamefont {Chan}},\ }\bibfield  {title} {\bibinfo {title} {Quantum embedding theories},\ }\href@noop {} {\bibfield  {journal} {\bibinfo  {journal} {Acc. Chem. Res.}\ }\textbf {\bibinfo {volume} {49}},\ \bibinfo {pages} {2705} (\bibinfo {year} {2016})}\BibitemShut {NoStop}%
\bibitem [{\citenamefont {Goodpaster}\ \emph {et~al.}(2014)\citenamefont {Goodpaster}, \citenamefont {Barnes}, \citenamefont {Manby},\ and\ \citenamefont {Miller}}]{Goodpaster2014}%
  \BibitemOpen
  \bibfield  {author} {\bibinfo {author} {\bibfnamefont {J.~D.}\ \bibnamefont {Goodpaster}}, \bibinfo {author} {\bibfnamefont {T.~A.}\ \bibnamefont {Barnes}}, \bibinfo {author} {\bibfnamefont {F.~R.}\ \bibnamefont {Manby}},\ and\ \bibinfo {author} {\bibfnamefont {T.~F.}\ \bibnamefont {Miller}},\ }\bibfield  {title} {\bibinfo {title} {Accurate and systematically improvable density functional theory embedding for correlated wavefunctions},\ }\href@noop {} {\bibfield  {journal} {\bibinfo  {journal} {J. Chem. Phys.}\ }\textbf {\bibinfo {volume} {140}},\ \bibinfo {pages} {140} (\bibinfo {year} {2014})}\BibitemShut {NoStop}%
\bibitem [{\citenamefont {Jacob}\ and\ \citenamefont {Neugebauer}(2014)}]{Jacob2014}%
  \BibitemOpen
  \bibfield  {author} {\bibinfo {author} {\bibfnamefont {C.~R.}\ \bibnamefont {Jacob}}\ and\ \bibinfo {author} {\bibfnamefont {J.}~\bibnamefont {Neugebauer}},\ }\bibfield  {title} {\bibinfo {title} {Subsystem density-functional theory},\ }\href@noop {} {\bibfield  {journal} {\bibinfo  {journal} {WIREs: Comput. Mol. Sci.}\ }\textbf {\bibinfo {volume} {4}},\ \bibinfo {pages} {325} (\bibinfo {year} {2014})}\BibitemShut {NoStop}%
\bibitem [{\citenamefont {Welborn}\ \emph {et~al.}(2016)\citenamefont {Welborn}, \citenamefont {Tsuchimochi},\ and\ \citenamefont {Van~Voorhis}}]{Welborn2016}%
  \BibitemOpen
  \bibfield  {author} {\bibinfo {author} {\bibfnamefont {M.}~\bibnamefont {Welborn}}, \bibinfo {author} {\bibfnamefont {T.}~\bibnamefont {Tsuchimochi}},\ and\ \bibinfo {author} {\bibfnamefont {T.}~\bibnamefont {Van~Voorhis}},\ }\bibfield  {title} {\bibinfo {title} {Bootstrap embedding: An internally consistent fragment-based method},\ }\href@noop {} {\bibfield  {journal} {\bibinfo  {journal} {J. Chem. Phys.}\ }\textbf {\bibinfo {volume} {145}},\ \bibinfo {pages} {074102} (\bibinfo {year} {2016})}\BibitemShut {NoStop}%
\bibitem [{\citenamefont {Libisch}\ \emph {et~al.}(2017)\citenamefont {Libisch}, \citenamefont {Marsman}, \citenamefont {Burgd\"orfer},\ and\ \citenamefont {Kresse}}]{Libisch2017}%
  \BibitemOpen
  \bibfield  {author} {\bibinfo {author} {\bibfnamefont {F.}~\bibnamefont {Libisch}}, \bibinfo {author} {\bibfnamefont {M.}~\bibnamefont {Marsman}}, \bibinfo {author} {\bibfnamefont {J.}~\bibnamefont {Burgd\"orfer}},\ and\ \bibinfo {author} {\bibfnamefont {G.}~\bibnamefont {Kresse}},\ }\bibfield  {title} {\bibinfo {title} {Embedding for bulk systems using localized atomic orbitals},\ }\href@noop {} {\bibfield  {journal} {\bibinfo  {journal} {J. Chem. Phys.}\ }\textbf {\bibinfo {volume} {147}},\ \bibinfo {pages} {034110} (\bibinfo {year} {2017})}\BibitemShut {NoStop}%
\bibitem [{\citenamefont {Chulhai}\ and\ \citenamefont {Goodpaster}(2018)}]{Chulhai2018}%
  \BibitemOpen
  \bibfield  {author} {\bibinfo {author} {\bibfnamefont {D.~V.}\ \bibnamefont {Chulhai}}\ and\ \bibinfo {author} {\bibfnamefont {J.~D.}\ \bibnamefont {Goodpaster}},\ }\bibfield  {title} {\bibinfo {title} {Projection-based correlated wave function in density functional theory embedding for periodic systems},\ }\href@noop {} {\bibfield  {journal} {\bibinfo  {journal} {J. Chem. Theory Comput.}\ }\textbf {\bibinfo {volume} {14}},\ \bibinfo {pages} {1928} (\bibinfo {year} {2018})}\BibitemShut {NoStop}%
\bibitem [{\citenamefont {Fertitta}\ and\ \citenamefont {Booth}(2018)}]{Fertitta2018}%
  \BibitemOpen
  \bibfield  {author} {\bibinfo {author} {\bibfnamefont {E.}~\bibnamefont {Fertitta}}\ and\ \bibinfo {author} {\bibfnamefont {G.~H.}\ \bibnamefont {Booth}},\ }\bibfield  {title} {\bibinfo {title} {Rigorous wave function embedding with dynamical fluctuations},\ }\href@noop {} {\bibfield  {journal} {\bibinfo  {journal} {Phys. Rev. B}\ }\textbf {\bibinfo {volume} {98}},\ \bibinfo {pages} {235132} (\bibinfo {year} {2018})}\BibitemShut {NoStop}%
\bibitem [{\citenamefont {Zhu}\ \emph {et~al.}(2019)\citenamefont {Zhu}, \citenamefont {Cui},\ and\ \citenamefont {Chan}}]{Zhu2019}%
  \BibitemOpen
  \bibfield  {author} {\bibinfo {author} {\bibfnamefont {T.}~\bibnamefont {Zhu}}, \bibinfo {author} {\bibfnamefont {Z.-H.}\ \bibnamefont {Cui}},\ and\ \bibinfo {author} {\bibfnamefont {G.~K.-L.}\ \bibnamefont {Chan}},\ }\bibfield  {title} {\bibinfo {title} {Efficient formulation of ab initio quantum embedding in periodic systems: Dynamical mean-field theory},\ }\href@noop {} {\bibfield  {journal} {\bibinfo  {journal} {J. Chem. Theory Comput.}\ }\textbf {\bibinfo {volume} {16}},\ \bibinfo {pages} {141} (\bibinfo {year} {2019})}\BibitemShut {NoStop}%
\bibitem [{\citenamefont {Lacombe}\ and\ \citenamefont {Maitra}(2020)}]{Lacombe2020}%
  \BibitemOpen
  \bibfield  {author} {\bibinfo {author} {\bibfnamefont {L.}~\bibnamefont {Lacombe}}\ and\ \bibinfo {author} {\bibfnamefont {N.~T.}\ \bibnamefont {Maitra}},\ }\bibfield  {title} {\bibinfo {title} {Embedding via the exact factorization approach},\ }\href@noop {} {\bibfield  {journal} {\bibinfo  {journal} {Phys. Rev. Lett.}\ }\textbf {\bibinfo {volume} {124}},\ \bibinfo {pages} {206401} (\bibinfo {year} {2020})}\BibitemShut {NoStop}%
\bibitem [{\citenamefont {Cui}\ \emph {et~al.}(2020)\citenamefont {Cui}, \citenamefont {Zhu},\ and\ \citenamefont {Chan}}]{Cui2020}%
  \BibitemOpen
  \bibfield  {author} {\bibinfo {author} {\bibfnamefont {Z.~H.}\ \bibnamefont {Cui}}, \bibinfo {author} {\bibfnamefont {T.}~\bibnamefont {Zhu}},\ and\ \bibinfo {author} {\bibfnamefont {G.~K.~L.}\ \bibnamefont {Chan}},\ }\bibfield  {title} {\bibinfo {title} {Efficient implementation of ab initio quantum embedding in periodic systems: Density matrix embedding theory},\ }\href@noop {} {\bibfield  {journal} {\bibinfo  {journal} {J. Chem. Theory Comput.}\ }\textbf {\bibinfo {volume} {16}},\ \bibinfo {pages} {119} (\bibinfo {year} {2020})}\BibitemShut {NoStop}%
\bibitem [{\citenamefont {Jones}\ \emph {et~al.}(2020)\citenamefont {Jones}, \citenamefont {Mosquera}, \citenamefont {Schatz},\ and\ \citenamefont {Ratner}}]{Jones2020}%
  \BibitemOpen
  \bibfield  {author} {\bibinfo {author} {\bibfnamefont {L.~O.}\ \bibnamefont {Jones}}, \bibinfo {author} {\bibfnamefont {M.~A.}\ \bibnamefont {Mosquera}}, \bibinfo {author} {\bibfnamefont {G.~C.}\ \bibnamefont {Schatz}},\ and\ \bibinfo {author} {\bibfnamefont {M.~A.}\ \bibnamefont {Ratner}},\ }\bibfield  {title} {\bibinfo {title} {Embedding methods for quantum chemistry: Applications from materials to life sciences},\ }\href@noop {} {\bibfield  {journal} {\bibinfo  {journal} {JACS}\ }\textbf {\bibinfo {volume} {142}},\ \bibinfo {pages} {3281} (\bibinfo {year} {2020})}\BibitemShut {NoStop}%
\bibitem [{\citenamefont {Pham}\ \emph {et~al.}(2020)\citenamefont {Pham}, \citenamefont {Hermes},\ and\ \citenamefont {Gagliardi}}]{Pham2020}%
  \BibitemOpen
  \bibfield  {author} {\bibinfo {author} {\bibfnamefont {H.~Q.}\ \bibnamefont {Pham}}, \bibinfo {author} {\bibfnamefont {M.~R.}\ \bibnamefont {Hermes}},\ and\ \bibinfo {author} {\bibfnamefont {L.}~\bibnamefont {Gagliardi}},\ }\bibfield  {title} {\bibinfo {title} {Periodic electronic structure calculations with the density matrix embedding theory},\ }\href@noop {} {\bibfield  {journal} {\bibinfo  {journal} {J. Chem. Theory Comput.}\ }\textbf {\bibinfo {volume} {16}},\ \bibinfo {pages} {130} (\bibinfo {year} {2020})}\BibitemShut {NoStop}%
\bibitem [{\citenamefont {Hégely}\ \emph {et~al.}(2016)\citenamefont {Hégely}, \citenamefont {Nagy}, \citenamefont {Ferenczy},\ and\ \citenamefont {Kállay}}]{Hegely}%
  \BibitemOpen
  \bibfield  {author} {\bibinfo {author} {\bibfnamefont {B.}~\bibnamefont {Hégely}}, \bibinfo {author} {\bibfnamefont {P.~R.}\ \bibnamefont {Nagy}}, \bibinfo {author} {\bibfnamefont {G.~G.}\ \bibnamefont {Ferenczy}},\ and\ \bibinfo {author} {\bibfnamefont {M.}~\bibnamefont {Kállay}},\ }\bibfield  {title} {\bibinfo {title} {Exact density functional and wave function embedding schemes based on orbital localization},\ }\href@noop {} {\bibfield  {journal} {\bibinfo  {journal} {J. Chem. Phys.}\ }\textbf {\bibinfo {volume} {145}},\ \bibinfo {pages} {064107} (\bibinfo {year} {2016})}\BibitemShut {NoStop}%
\bibitem [{\citenamefont {Ma}\ \emph {et~al.}(2021)\citenamefont {Ma}, \citenamefont {Sheng}, \citenamefont {Govoni},\ and\ \citenamefont {Galli}}]{Ma2021}%
  \BibitemOpen
  \bibfield  {author} {\bibinfo {author} {\bibfnamefont {H.}~\bibnamefont {Ma}}, \bibinfo {author} {\bibfnamefont {N.}~\bibnamefont {Sheng}}, \bibinfo {author} {\bibfnamefont {M.}~\bibnamefont {Govoni}},\ and\ \bibinfo {author} {\bibfnamefont {G.}~\bibnamefont {Galli}},\ }\bibfield  {title} {\bibinfo {title} {Quantum embedding theory for strongly correlated states in materials},\ }\href@noop {} {\bibfield  {journal} {\bibinfo  {journal} {J. Chem. Theory Comput.}\ }\textbf {\bibinfo {volume} {17}},\ \bibinfo {pages} {2116} (\bibinfo {year} {2021})}\BibitemShut {NoStop}%
\bibitem [{\citenamefont {Wachter-Lehn}\ \emph {et~al.}(2022)\citenamefont {Wachter-Lehn}, \citenamefont {Fink},\ and\ \citenamefont {H\"{o}fener}}]{WachterLehn2022}%
  \BibitemOpen
  \bibfield  {author} {\bibinfo {author} {\bibfnamefont {M.~T.}\ \bibnamefont {Wachter-Lehn}}, \bibinfo {author} {\bibfnamefont {K.}~\bibnamefont {Fink}},\ and\ \bibinfo {author} {\bibfnamefont {S.}~\bibnamefont {H\"{o}fener}},\ }\bibfield  {title} {\bibinfo {title} {Wavefunction frozen-density embedding with one-dimensional periodicity: Electronic polarization effects from local perturbations},\ }\href@noop {} {\bibfield  {journal} {\bibinfo  {journal} {J. Chem. Phys.}\ }\textbf {\bibinfo {volume} {157}},\ \bibinfo {pages} {134109} (\bibinfo {year} {2022})}\BibitemShut {NoStop}%
\bibitem [{\citenamefont {Nusspickel}\ and\ \citenamefont {Booth}(2022)}]{Nusspickel2022}%
  \BibitemOpen
  \bibfield  {author} {\bibinfo {author} {\bibfnamefont {M.}~\bibnamefont {Nusspickel}}\ and\ \bibinfo {author} {\bibfnamefont {G.~H.}\ \bibnamefont {Booth}},\ }\bibfield  {title} {\bibinfo {title} {Systematic improvability in quantum embedding for real materials},\ }\href@noop {} {\bibfield  {journal} {\bibinfo  {journal} {Phys. Rev. X}\ }\textbf {\bibinfo {volume} {12}},\ \bibinfo {pages} {011046} (\bibinfo {year} {2022})}\BibitemShut {NoStop}%
\bibitem [{\citenamefont {Birkenheuer}\ \emph {et~al.}(2006)\citenamefont {Birkenheuer}, \citenamefont {Fulde},\ and\ \citenamefont {Stoll}}]{Birkenheuer2006}%
  \BibitemOpen
  \bibfield  {author} {\bibinfo {author} {\bibfnamefont {U.}~\bibnamefont {Birkenheuer}}, \bibinfo {author} {\bibfnamefont {P.}~\bibnamefont {Fulde}},\ and\ \bibinfo {author} {\bibfnamefont {H.}~\bibnamefont {Stoll}},\ }\bibfield  {title} {\bibinfo {title} {A simplified method for the computation of correlation effects on the band structure of semiconductors},\ }\href@noop {} {\bibfield  {journal} {\bibinfo  {journal} {Theor. Chem. Acc.}\ }\textbf {\bibinfo {volume} {116}},\ \bibinfo {pages} {398} (\bibinfo {year} {2006})}\BibitemShut {NoStop}%
\bibitem [{\citenamefont {Hozoi}\ \emph {et~al.}(2009)\citenamefont {Hozoi}, \citenamefont {Birkenheuer}, \citenamefont {Stoll},\ and\ \citenamefont {Fulde}}]{Hozoi2009}%
  \BibitemOpen
  \bibfield  {author} {\bibinfo {author} {\bibfnamefont {L.}~\bibnamefont {Hozoi}}, \bibinfo {author} {\bibfnamefont {U.}~\bibnamefont {Birkenheuer}}, \bibinfo {author} {\bibfnamefont {H.}~\bibnamefont {Stoll}},\ and\ \bibinfo {author} {\bibfnamefont {P.}~\bibnamefont {Fulde}},\ }\bibfield  {title} {\bibinfo {title} {Spin-state transition and spin-polaron physics in cobalt oxide perovskites:ab initioapproach based on quantum chemical methods},\ }\href@noop {} {\bibfield  {journal} {\bibinfo  {journal} {New J. Phys.}\ }\textbf {\bibinfo {volume} {11}},\ \bibinfo {pages} {023023} (\bibinfo {year} {2009})}\BibitemShut {NoStop}%
\bibitem [{\citenamefont {de~Lara-Castells}\ and\ \citenamefont {Mitrushchenkov}(2011)}]{de_lara-castells2011}%
  \BibitemOpen
  \bibfield  {author} {\bibinfo {author} {\bibfnamefont {M.~P.}\ \bibnamefont {de~Lara-Castells}}\ and\ \bibinfo {author} {\bibfnamefont {A.~O.}\ \bibnamefont {Mitrushchenkov}},\ }\bibfield  {title} {\bibinfo {title} {A {Finite} {Cluster} {Approach} to an {Extended} {Transition} {Metal} {Oxide}: {A} {Wave} {Function} {Based} {Study}},\ }\href {https://doi.org/10.1021/jp203654m} {\bibfield  {journal} {\bibinfo  {journal} {J. Phys. Chem. C}\ }\textbf {\bibinfo {volume} {115}},\ \bibinfo {pages} {17540} (\bibinfo {year} {2011})}\BibitemShut {NoStop}%
\bibitem [{\citenamefont {Stoyanova}\ \emph {et~al.}(2014)\citenamefont {Stoyanova}, \citenamefont {Mitrushchenkov}, \citenamefont {Hozoi}, \citenamefont {Stoll},\ and\ \citenamefont {Fulde}}]{Stoyanova2014}%
  \BibitemOpen
  \bibfield  {author} {\bibinfo {author} {\bibfnamefont {A.}~\bibnamefont {Stoyanova}}, \bibinfo {author} {\bibfnamefont {A.~O.}\ \bibnamefont {Mitrushchenkov}}, \bibinfo {author} {\bibfnamefont {L.}~\bibnamefont {Hozoi}}, \bibinfo {author} {\bibfnamefont {H.}~\bibnamefont {Stoll}},\ and\ \bibinfo {author} {\bibfnamefont {P.}~\bibnamefont {Fulde}},\ }\bibfield  {title} {\bibinfo {title} {Electron correlation effects in diamond: A wave-function quantum-chemistry study of the quasiparticle band structure},\ }\href@noop {} {\bibfield  {journal} {\bibinfo  {journal} {Phys. Rev. B}\ }\textbf {\bibinfo {volume} {89}},\ \bibinfo {pages} {235121} (\bibinfo {year} {2014})}\BibitemShut {NoStop}%
\bibitem [{\citenamefont {Masur}\ \emph {et~al.}(2016)\citenamefont {Masur}, \citenamefont {Sch\"utz}, \citenamefont {Maschio},\ and\ \citenamefont {Usvyat}}]{masur2016}%
  \BibitemOpen
  \bibfield  {author} {\bibinfo {author} {\bibfnamefont {O.}~\bibnamefont {Masur}}, \bibinfo {author} {\bibfnamefont {M.}~\bibnamefont {Sch\"utz}}, \bibinfo {author} {\bibfnamefont {L.}~\bibnamefont {Maschio}},\ and\ \bibinfo {author} {\bibfnamefont {D.}~\bibnamefont {Usvyat}},\ }\bibfield  {title} {\bibinfo {title} {Fragment-based direct-local-ring-coupled-cluster doubles treatment embedded in the periodic hartree–fock solution},\ }\href@noop {} {\bibfield  {journal} {\bibinfo  {journal} {J. Chem. Theory Comput.}\ }\textbf {\bibinfo {volume} {12}},\ \bibinfo {pages} {5145} (\bibinfo {year} {2016})}\BibitemShut {NoStop}%
\bibitem [{\citenamefont {Usvyat}\ \emph {et~al.}(2018)\citenamefont {Usvyat}, \citenamefont {Maschio},\ and\ \citenamefont {Sch{\"u}tz}}]{usvyat18}%
  \BibitemOpen
  \bibfield  {author} {\bibinfo {author} {\bibfnamefont {D.}~\bibnamefont {Usvyat}}, \bibinfo {author} {\bibfnamefont {L.}~\bibnamefont {Maschio}},\ and\ \bibinfo {author} {\bibfnamefont {M.}~\bibnamefont {Sch{\"u}tz}},\ }\bibfield  {title} {\bibinfo {title} {Periodic and fragment models based on the local correlation approach},\ }\href@noop {} {\bibfield  {journal} {\bibinfo  {journal} {WIREs: Comput. Mol. Sci.}\ }\textbf {\bibinfo {volume} {8}},\ \bibinfo {pages} {e1357} (\bibinfo {year} {2018})}\BibitemShut {NoStop}%
\bibitem [{\citenamefont {Lin}\ \emph {et~al.}(2020)\citenamefont {Lin}, \citenamefont {Maschio}, \citenamefont {Kats}, \citenamefont {Usvyat},\ and\ \citenamefont {Heine}}]{usvyat20}%
  \BibitemOpen
  \bibfield  {author} {\bibinfo {author} {\bibfnamefont {H.~H.}\ \bibnamefont {Lin}}, \bibinfo {author} {\bibfnamefont {L.}~\bibnamefont {Maschio}}, \bibinfo {author} {\bibfnamefont {D.}~\bibnamefont {Kats}}, \bibinfo {author} {\bibfnamefont {D.}~\bibnamefont {Usvyat}},\ and\ \bibinfo {author} {\bibfnamefont {T.}~\bibnamefont {Heine}},\ }\bibfield  {title} {\bibinfo {title} {Fragment-based restricted active space configuration interaction with second-order corrections embedded in periodic hartree–fock wave function},\ }\href@noop {} {\bibfield  {journal} {\bibinfo  {journal} {J. Chem. Theory Comput.}\ }\textbf {\bibinfo {volume} {16}},\ \bibinfo {pages} {7100} (\bibinfo {year} {2020})}\BibitemShut {NoStop}%
\bibitem [{\citenamefont {Sch{\"a}fer}\ \emph {et~al.}(2021)\citenamefont {Sch{\"a}fer}, \citenamefont {Libisch}, \citenamefont {Kresse},\ and\ \citenamefont {Gr{\"u}neis}}]{schaefer21}%
  \BibitemOpen
  \bibfield  {author} {\bibinfo {author} {\bibfnamefont {T.}~\bibnamefont {Sch{\"a}fer}}, \bibinfo {author} {\bibfnamefont {F.}~\bibnamefont {Libisch}}, \bibinfo {author} {\bibfnamefont {G.}~\bibnamefont {Kresse}},\ and\ \bibinfo {author} {\bibfnamefont {A.}~\bibnamefont {Gr{\"u}neis}},\ }\bibfield  {title} {\bibinfo {title} {Local embedding of coupled cluster theory into the random phase approximation using plane waves},\ }\href@noop {} {\bibfield  {journal} {\bibinfo  {journal} {J. Chem. Phys.}\ }\textbf {\bibinfo {volume} {154}},\ \bibinfo {pages} {011101} (\bibinfo {year} {2021})}\BibitemShut {NoStop}%
\bibitem [{\citenamefont {Christlmaier}\ \emph {et~al.}(2022)\citenamefont {Christlmaier}, \citenamefont {Kats}, \citenamefont {Alavi},\ and\ \citenamefont {Usvyat}}]{christlmaier21}%
  \BibitemOpen
  \bibfield  {author} {\bibinfo {author} {\bibfnamefont {E.~M.}\ \bibnamefont {Christlmaier}}, \bibinfo {author} {\bibfnamefont {D.}~\bibnamefont {Kats}}, \bibinfo {author} {\bibfnamefont {A.}~\bibnamefont {Alavi}},\ and\ \bibinfo {author} {\bibfnamefont {D.}~\bibnamefont {Usvyat}},\ }\bibfield  {title} {\bibinfo {title} {Full configuration interaction quantum monte carlo treatment of fragments embedded in a periodic mean field},\ }\href@noop {} {\bibfield  {journal} {\bibinfo  {journal} {J. Chem. Phys.}\ }\textbf {\bibinfo {volume} {156}},\ \bibinfo {pages} {074109} (\bibinfo {year} {2022})}\BibitemShut {NoStop}%
\bibitem [{\citenamefont {Lau}\ \emph {et~al.}(2021)\citenamefont {Lau}, \citenamefont {Knizia},\ and\ \citenamefont {Berkelbach}}]{Berkelbach21}%
  \BibitemOpen
  \bibfield  {author} {\bibinfo {author} {\bibfnamefont {B.~T.~G.}\ \bibnamefont {Lau}}, \bibinfo {author} {\bibfnamefont {G.}~\bibnamefont {Knizia}},\ and\ \bibinfo {author} {\bibfnamefont {T.~C.}\ \bibnamefont {Berkelbach}},\ }\bibfield  {title} {\bibinfo {title} {Regional embedding enables high-level quantum chemistry for surface science},\ }\href@noop {} {\bibfield  {journal} {\bibinfo  {journal} {J. Phys. Chem. Lett.}\ }\textbf {\bibinfo {volume} {12}},\ \bibinfo {pages} {1104} (\bibinfo {year} {2021})}\BibitemShut {NoStop}%
\bibitem [{\citenamefont {Burow}\ \emph {et~al.}(2009)\citenamefont {Burow}, \citenamefont {Sierka}, \citenamefont {D\"{o}bler},\ and\ \citenamefont {Sauer}}]{burow2009}%
  \BibitemOpen
  \bibfield  {author} {\bibinfo {author} {\bibfnamefont {A.~M.}\ \bibnamefont {Burow}}, \bibinfo {author} {\bibfnamefont {M.}~\bibnamefont {Sierka}}, \bibinfo {author} {\bibfnamefont {J.}~\bibnamefont {D\"{o}bler}},\ and\ \bibinfo {author} {\bibfnamefont {J.}~\bibnamefont {Sauer}},\ }\bibfield  {title} {\bibinfo {title} {Point defects in {CaF2} and {CeO2} investigated by the periodic electrostatic embedded cluster method},\ }\href {https://doi.org/10.1063/1.3123527} {\bibfield  {journal} {\bibinfo  {journal} {J. Chem. Phys.}\ }\textbf {\bibinfo {volume} {130}},\ \bibinfo {pages} {174710} (\bibinfo {year} {2009})}\BibitemShut {NoStop}%
\bibitem [{\citenamefont {Wachters}\ and\ \citenamefont {Nieuwpoort}(1972)}]{Wachters1972}%
  \BibitemOpen
  \bibfield  {author} {\bibinfo {author} {\bibfnamefont {A.~J.~H.}\ \bibnamefont {Wachters}}\ and\ \bibinfo {author} {\bibfnamefont {W.~C.}\ \bibnamefont {Nieuwpoort}},\ }\bibfield  {title} {\bibinfo {title} {Ab initio calculations on {KN}i${\mathrm{f}}_{3}$: Ligand-field effects},\ }\href {https://doi.org/10.1103/PhysRevB.5.4291} {\bibfield  {journal} {\bibinfo  {journal} {Phys. Rev. B}\ }\textbf {\bibinfo {volume} {5}},\ \bibinfo {pages} {4291} (\bibinfo {year} {1972})}\BibitemShut {NoStop}%
\bibitem [{\citenamefont {Sousa}\ \emph {et~al.}(1993)\citenamefont {Sousa}, \citenamefont {Casanovas}, \citenamefont {Rubio},\ and\ \citenamefont {Illas}}]{sousa1993}%
  \BibitemOpen
  \bibfield  {author} {\bibinfo {author} {\bibfnamefont {C.}~\bibnamefont {Sousa}}, \bibinfo {author} {\bibfnamefont {J.}~\bibnamefont {Casanovas}}, \bibinfo {author} {\bibfnamefont {J.}~\bibnamefont {Rubio}},\ and\ \bibinfo {author} {\bibfnamefont {F.}~\bibnamefont {Illas}},\ }\bibfield  {title} {\bibinfo {title} {Madelung fields from optimized point charges for ab initio cluster model calculations on ionic systems},\ }\href {https://doi.org/10.1002/jcc.540140608} {\bibfield  {journal} {\bibinfo  {journal} {J. Comp. Chem.}\ }\textbf {\bibinfo {volume} {14}},\ \bibinfo {pages} {680} (\bibinfo {year} {1993})}\BibitemShut {NoStop}%
\bibitem [{\citenamefont {Klintenberg}\ \emph {et~al.}(2000)\citenamefont {Klintenberg}, \citenamefont {Derenzo},\ and\ \citenamefont {Weber}}]{Klintenberg2000}%
  \BibitemOpen
  \bibfield  {author} {\bibinfo {author} {\bibfnamefont {M.}~\bibnamefont {Klintenberg}}, \bibinfo {author} {\bibfnamefont {S.}~\bibnamefont {Derenzo}},\ and\ \bibinfo {author} {\bibfnamefont {M.}~\bibnamefont {Weber}},\ }\bibfield  {title} {\bibinfo {title} {Accurate crystal fields for embedded cluster calculations},\ }\href {https://doi.org/10.1016/s0010-4655(00)00071-0} {\bibfield  {journal} {\bibinfo  {journal} {Computer Physics Communications}\ }\textbf {\bibinfo {volume} {131}},\ \bibinfo {pages} {120–128} (\bibinfo {year} {2000})}\BibitemShut {NoStop}%
\bibitem [{\citenamefont {Gell\'{e}}\ and\ \citenamefont {Lepetit}(2008)}]{gelle2008}%
  \BibitemOpen
  \bibfield  {author} {\bibinfo {author} {\bibfnamefont {A.}~\bibnamefont {Gell\'{e}}}\ and\ \bibinfo {author} {\bibfnamefont {M.-B.}\ \bibnamefont {Lepetit}},\ }\bibfield  {title} {\bibinfo {title} {Fast calculation of the electrostatic potential in ionic crystals by direct summation method},\ }\href {https://doi.org/10.1063/1.2931458} {\bibfield  {journal} {\bibinfo  {journal} {J. Chem. Phys.}\ }\textbf {\bibinfo {volume} {128}},\ \bibinfo {pages} {244716} (\bibinfo {year} {2008})}\BibitemShut {NoStop}%
\bibitem [{\citenamefont {Sushko}\ and\ \citenamefont {Abarenkov}(2010)}]{sushko2010}%
  \BibitemOpen
  \bibfield  {author} {\bibinfo {author} {\bibfnamefont {P.~V.}\ \bibnamefont {Sushko}}\ and\ \bibinfo {author} {\bibfnamefont {I.~V.}\ \bibnamefont {Abarenkov}},\ }\bibfield  {title} {\bibinfo {title} {General {Purpose} {Electrostatic} {Embedding} {Potential}},\ }\href {https://doi.org/10.1021/ct900480p} {\bibfield  {journal} {\bibinfo  {journal} {J. Chem. Theory Comp.}\ }\textbf {\bibinfo {volume} {6}},\ \bibinfo {pages} {1323} (\bibinfo {year} {2010})}\BibitemShut {NoStop}%
\bibitem [{\citenamefont {Illas}\ \emph {et~al.}(1993)\citenamefont {Illas}, \citenamefont {Casanovas}, \citenamefont {Garc\'{\i}a-Bach}, \citenamefont {Caballol},\ and\ \citenamefont {Castell}}]{Illas1993}%
  \BibitemOpen
  \bibfield  {author} {\bibinfo {author} {\bibfnamefont {F.}~\bibnamefont {Illas}}, \bibinfo {author} {\bibfnamefont {J.}~\bibnamefont {Casanovas}}, \bibinfo {author} {\bibfnamefont {M.~A.}\ \bibnamefont {Garc\'{\i}a-Bach}}, \bibinfo {author} {\bibfnamefont {R.}~\bibnamefont {Caballol}},\ and\ \bibinfo {author} {\bibfnamefont {O.}~\bibnamefont {Castell}},\ }\bibfield  {title} {\bibinfo {title} {Towards an ab initio description of magnetism in ionic solids},\ }\href {https://doi.org/10.1103/PhysRevLett.71.3549} {\bibfield  {journal} {\bibinfo  {journal} {Phys. Rev. Lett.}\ }\textbf {\bibinfo {volume} {71}},\ \bibinfo {pages} {3549} (\bibinfo {year} {1993})}\BibitemShut {NoStop}%
\bibitem [{\citenamefont {de~Graaf}\ and\ \citenamefont {Illas}(2000)}]{deGraaf2000}%
  \BibitemOpen
  \bibfield  {author} {\bibinfo {author} {\bibfnamefont {C.}~\bibnamefont {de~Graaf}}\ and\ \bibinfo {author} {\bibfnamefont {F.}~\bibnamefont {Illas}},\ }\bibfield  {title} {\bibinfo {title} {Electronic structure and magnetic interactions of the spin-chain compounds ${\mathrm{ca}}_{2}{\mathrm{cuo}}_{3}$ and ${\mathrm{sr}}_{2}{\mathrm{cuo}}_{3}$},\ }\href {https://doi.org/10.1103/PhysRevB.63.014404} {\bibfield  {journal} {\bibinfo  {journal} {Phys. Rev. B}\ }\textbf {\bibinfo {volume} {63}},\ \bibinfo {pages} {014404} (\bibinfo {year} {2000})}\BibitemShut {NoStop}%
\bibitem [{\citenamefont {Sousa}\ and\ \citenamefont {Illas}(2001)}]{Sousa2001}%
  \BibitemOpen
  \bibfield  {author} {\bibinfo {author} {\bibfnamefont {C.}~\bibnamefont {Sousa}}\ and\ \bibinfo {author} {\bibfnamefont {F.}~\bibnamefont {Illas}},\ }\bibfield  {title} {\bibinfo {title} {On the accurate prediction of the optical absorption energy of f-centers in mgo from explicitly correlated ab initio cluster model calculations},\ }\href@noop {} {\bibfield  {journal} {\bibinfo  {journal} {J. Chem. Phys.}\ }\textbf {\bibinfo {volume} {115}},\ \bibinfo {pages} {1435} (\bibinfo {year} {2001})}\BibitemShut {NoStop}%
\bibitem [{\citenamefont {Pradipto}\ \emph {et~al.}(2012)\citenamefont {Pradipto}, \citenamefont {Maurice}, \citenamefont {Guih\'ery}, \citenamefont {de~Graaf},\ and\ \citenamefont {Broer}}]{Pradipto2012}%
  \BibitemOpen
  \bibfield  {author} {\bibinfo {author} {\bibfnamefont {A.-M.}\ \bibnamefont {Pradipto}}, \bibinfo {author} {\bibfnamefont {R.}~\bibnamefont {Maurice}}, \bibinfo {author} {\bibfnamefont {N.}~\bibnamefont {Guih\'ery}}, \bibinfo {author} {\bibfnamefont {C.}~\bibnamefont {de~Graaf}},\ and\ \bibinfo {author} {\bibfnamefont {R.}~\bibnamefont {Broer}},\ }\bibfield  {title} {\bibinfo {title} {First-principles study of magnetic interactions in cupric oxide},\ }\href {https://doi.org/10.1103/PhysRevB.85.014409} {\bibfield  {journal} {\bibinfo  {journal} {Phys. Rev. B}\ }\textbf {\bibinfo {volume} {85}},\ \bibinfo {pages} {014409} (\bibinfo {year} {2012})}\BibitemShut {NoStop}%
\bibitem [{\citenamefont {Domingo}\ \emph {et~al.}(2012)\citenamefont {Domingo}, \citenamefont {Rodr{\'\i}guez-Fortea}, \citenamefont {Swart}, \citenamefont {de~Graaf},\ and\ \citenamefont {Broer}}]{Domingo2012-jb}%
  \BibitemOpen
  \bibfield  {author} {\bibinfo {author} {\bibfnamefont {A.}~\bibnamefont {Domingo}}, \bibinfo {author} {\bibfnamefont {A.}~\bibnamefont {Rodr{\'\i}guez-Fortea}}, \bibinfo {author} {\bibfnamefont {M.}~\bibnamefont {Swart}}, \bibinfo {author} {\bibfnamefont {C.}~\bibnamefont {de~Graaf}},\ and\ \bibinfo {author} {\bibfnamefont {R.}~\bibnamefont {Broer}},\ }\bibfield  {title} {\bibinfo {title} {Ab initioabsorption spectrum of {NiO} combining molecular dynamics with the embedded cluster approach in a discrete reaction field},\ }\href@noop {} {\bibfield  {journal} {\bibinfo  {journal} {Phys. Rev. B}\ }\textbf {\bibinfo {volume} {85}},\ \bibinfo {pages} {155143} (\bibinfo {year} {2012})}\BibitemShut {NoStop}%
\bibitem [{\citenamefont {Kubas}\ \emph {et~al.}(2016)\citenamefont {Kubas}, \citenamefont {Berger}, \citenamefont {Oberhofer}, \citenamefont {Maganas}, \citenamefont {Reuter},\ and\ \citenamefont {Neese}}]{Kubas2016}%
  \BibitemOpen
  \bibfield  {author} {\bibinfo {author} {\bibfnamefont {A.}~\bibnamefont {Kubas}}, \bibinfo {author} {\bibfnamefont {D.}~\bibnamefont {Berger}}, \bibinfo {author} {\bibfnamefont {H.}~\bibnamefont {Oberhofer}}, \bibinfo {author} {\bibfnamefont {D.}~\bibnamefont {Maganas}}, \bibinfo {author} {\bibfnamefont {K.}~\bibnamefont {Reuter}},\ and\ \bibinfo {author} {\bibfnamefont {F.}~\bibnamefont {Neese}},\ }\bibfield  {title} {\bibinfo {title} {Surface adsorption energetics studied with “gold standard” wave-function-based ab initio methods: Small-molecule binding to tio2(110)},\ }\href {https://doi.org/10.1021/acs.jpclett.6b01845} {\bibfield  {journal} {\bibinfo  {journal} {J. Phys. Chem. Lett.}\ }\textbf {\bibinfo {volume} {7}},\ \bibinfo {pages} {4207} (\bibinfo {year} {2016})}\BibitemShut {NoStop}%
\bibitem [{\citenamefont {Katukuri}\ \emph {et~al.}(2020)\citenamefont {Katukuri}, \citenamefont {Bogdanov}, \citenamefont {Weser}, \citenamefont {van~den Brink},\ and\ \citenamefont {Alavi}}]{Katukuri2020}%
  \BibitemOpen
  \bibfield  {author} {\bibinfo {author} {\bibfnamefont {V.~M.}\ \bibnamefont {Katukuri}}, \bibinfo {author} {\bibfnamefont {N.~A.}\ \bibnamefont {Bogdanov}}, \bibinfo {author} {\bibfnamefont {O.}~\bibnamefont {Weser}}, \bibinfo {author} {\bibfnamefont {J.}~\bibnamefont {van~den Brink}},\ and\ \bibinfo {author} {\bibfnamefont {A.}~\bibnamefont {Alavi}},\ }\bibfield  {title} {\bibinfo {title} {Electronic correlations and magnetic interactions in infinite-layer ndnio2},\ }\href@noop {} {\bibfield  {journal} {\bibinfo  {journal} {Phys. Rev. B}\ }\textbf {\bibinfo {volume} {102}},\ \bibinfo {pages} {241112(R)} (\bibinfo {year} {2020})}\BibitemShut {NoStop}%
\bibitem [{\citenamefont {Chen}\ \emph {et~al.}(2020)\citenamefont {Chen}, \citenamefont {Bogdanov}, \citenamefont {Usvyat}, \citenamefont {Fang}, \citenamefont {Michaelides},\ and\ \citenamefont {Alavi}}]{chen20}%
  \BibitemOpen
  \bibfield  {author} {\bibinfo {author} {\bibfnamefont {J.}~\bibnamefont {Chen}}, \bibinfo {author} {\bibfnamefont {N.~A.}\ \bibnamefont {Bogdanov}}, \bibinfo {author} {\bibfnamefont {D.}~\bibnamefont {Usvyat}}, \bibinfo {author} {\bibfnamefont {W.}~\bibnamefont {Fang}}, \bibinfo {author} {\bibfnamefont {A.}~\bibnamefont {Michaelides}},\ and\ \bibinfo {author} {\bibfnamefont {A.}~\bibnamefont {Alavi}},\ }\bibfield  {title} {\bibinfo {title} {The color center singlet state of oxygen vacancies in tio2},\ }\href@noop {} {\bibfield  {journal} {\bibinfo  {journal} {J. Chem. Phys.}\ }\textbf {\bibinfo {volume} {153}},\ \bibinfo {pages} {204704} (\bibinfo {year} {2020})}\BibitemShut {NoStop}%
\bibitem [{\citenamefont {Bogdanov}\ \emph {et~al.}(2021)\citenamefont {Bogdanov}, \citenamefont {Li~Manni}, \citenamefont {Sharma}, \citenamefont {Gunnarsson},\ and\ \citenamefont {Alavi}}]{Bogdanov2021}%
  \BibitemOpen
  \bibfield  {author} {\bibinfo {author} {\bibfnamefont {N.~A.}\ \bibnamefont {Bogdanov}}, \bibinfo {author} {\bibfnamefont {G.}~\bibnamefont {Li~Manni}}, \bibinfo {author} {\bibfnamefont {S.}~\bibnamefont {Sharma}}, \bibinfo {author} {\bibfnamefont {O.}~\bibnamefont {Gunnarsson}},\ and\ \bibinfo {author} {\bibfnamefont {A.}~\bibnamefont {Alavi}},\ }\bibfield  {title} {\bibinfo {title} {Enhancement of superexchange due to synergetic breathing and hopping in corner-sharing cuprates},\ }\href@noop {} {\bibfield  {journal} {\bibinfo  {journal} {Nature Physics}\ }\textbf {\bibinfo {volume} {18}},\ \bibinfo {pages} {190} (\bibinfo {year} {2021})}\BibitemShut {NoStop}%
\bibitem [{\citenamefont {Bhattacharyya}\ and\ \citenamefont {Hozoi}(2022)}]{Bhattacharyya2022}%
  \BibitemOpen
  \bibfield  {author} {\bibinfo {author} {\bibfnamefont {P.}~\bibnamefont {Bhattacharyya}}\ and\ \bibinfo {author} {\bibfnamefont {L.}~\bibnamefont {Hozoi}},\ }\bibfield  {title} {\bibinfo {title} {Yb3+ f-f excitations in naybse2 : Benchmarking embedded-cluster quantum chemical schemes for 4f insulators},\ }\href@noop {} {\bibfield  {journal} {\bibinfo  {journal} {Phys. Rev. B}\ }\textbf {\bibinfo {volume} {105}},\ \bibinfo {pages} {235117} (\bibinfo {year} {2022})}\BibitemShut {NoStop}%
\bibitem [{\citenamefont {Petersen}\ \emph {et~al.}(2022)\citenamefont {Petersen}, \citenamefont {R\"{o}ßler},\ and\ \citenamefont {Hozoi}}]{Petersen2022}%
  \BibitemOpen
  \bibfield  {author} {\bibinfo {author} {\bibfnamefont {T.}~\bibnamefont {Petersen}}, \bibinfo {author} {\bibfnamefont {U.~K.}\ \bibnamefont {R\"{o}ßler}},\ and\ \bibinfo {author} {\bibfnamefont {L.}~\bibnamefont {Hozoi}},\ }\bibfield  {title} {\bibinfo {title} {Quantum chemical insights into hexaboride electronic structures: correlations within the boron p-orbital subsystem},\ }\href@noop {} {\bibfield  {journal} {\bibinfo  {journal} {Commun. Phys.}\ }\textbf {\bibinfo {volume} {5}},\ \bibinfo {pages} {214} (\bibinfo {year} {2022})}\BibitemShut {NoStop}%
\bibitem [{\citenamefont {Nabi}\ \emph {et~al.}(2023)\citenamefont {Nabi}, \citenamefont {Staab}, \citenamefont {Mattioni}, \citenamefont {Kragskow}, \citenamefont {Reta}, \citenamefont {Skelton},\ and\ \citenamefont {Chilton}}]{Nabi2023}%
  \BibitemOpen
  \bibfield  {author} {\bibinfo {author} {\bibfnamefont {R.}~\bibnamefont {Nabi}}, \bibinfo {author} {\bibfnamefont {J.~K.}\ \bibnamefont {Staab}}, \bibinfo {author} {\bibfnamefont {A.}~\bibnamefont {Mattioni}}, \bibinfo {author} {\bibfnamefont {J.~G.~C.}\ \bibnamefont {Kragskow}}, \bibinfo {author} {\bibfnamefont {D.}~\bibnamefont {Reta}}, \bibinfo {author} {\bibfnamefont {J.~M.}\ \bibnamefont {Skelton}},\ and\ \bibinfo {author} {\bibfnamefont {N.~F.}\ \bibnamefont {Chilton}},\ }\bibfield  {title} {\bibinfo {title} {Accurate and efficient spin–phonon coupling and spin dynamics calculations for molecular solids},\ }\href@noop {} {\bibfield  {journal} {\bibinfo  {journal} {JACS}\ }\textbf {\bibinfo {volume} {45}},\ \bibinfo {pages} {24558} (\bibinfo {year} {2023})}\BibitemShut {NoStop}%
\bibitem [{\citenamefont {Barandiar\'{a}n}\ and\ \citenamefont {Seijo}(1988)}]{barandiaran1988}%
  \BibitemOpen
  \bibfield  {author} {\bibinfo {author} {\bibfnamefont {Z.}~\bibnamefont {Barandiar\'{a}n}}\ and\ \bibinfo {author} {\bibfnamefont {L.}~\bibnamefont {Seijo}},\ }\bibfield  {title} {\bibinfo {title} {The ab initio model potential representation of the crystalline environment. {Theoretical} study of the local distortion on {NaCl}:{Cu}+},\ }\href {https://doi.org/10.1063/1.455549} {\bibfield  {journal} {\bibinfo  {journal} {J. Chem. Phys.}\ }\textbf {\bibinfo {volume} {89}},\ \bibinfo {pages} {5739} (\bibinfo {year} {1988})}\BibitemShut {NoStop}%
\bibitem [{\citenamefont {Lepetit}\ \emph {et~al.}(2003)\citenamefont {Lepetit}, \citenamefont {Suaud}, \citenamefont {Gelle},\ and\ \citenamefont {Robert}}]{lepetit2003}%
  \BibitemOpen
  \bibfield  {author} {\bibinfo {author} {\bibfnamefont {M.-B.}\ \bibnamefont {Lepetit}}, \bibinfo {author} {\bibfnamefont {N.}~\bibnamefont {Suaud}}, \bibinfo {author} {\bibfnamefont {A.}~\bibnamefont {Gelle}},\ and\ \bibinfo {author} {\bibfnamefont {V.}~\bibnamefont {Robert}},\ }\bibfield  {title} {\bibinfo {title} {Environment effects on effective magnetic exchange integrals and local spectroscopy of extended strongly correlated systems},\ }\href {https://doi.org/10.1063/1.1540620} {\bibfield  {journal} {\bibinfo  {journal} {J. Chem. Phys.}\ }\textbf {\bibinfo {volume} {118}},\ \bibinfo {pages} {3966} (\bibinfo {year} {2003})}\BibitemShut {NoStop}%
\bibitem [{\citenamefont {Swerts}\ \emph {et~al.}(2008)\citenamefont {Swerts}, \citenamefont {Chibotaru}, \citenamefont {Lindh}, \citenamefont {Seijo}, \citenamefont {Barandiaran}, \citenamefont {Clima}, \citenamefont {Pierloot},\ and\ \citenamefont {Hendrickx}}]{swerts2008}%
  \BibitemOpen
  \bibfield  {author} {\bibinfo {author} {\bibfnamefont {B.}~\bibnamefont {Swerts}}, \bibinfo {author} {\bibfnamefont {L.~F.}\ \bibnamefont {Chibotaru}}, \bibinfo {author} {\bibfnamefont {R.}~\bibnamefont {Lindh}}, \bibinfo {author} {\bibfnamefont {L.}~\bibnamefont {Seijo}}, \bibinfo {author} {\bibfnamefont {Z.}~\bibnamefont {Barandiaran}}, \bibinfo {author} {\bibfnamefont {S.}~\bibnamefont {Clima}}, \bibinfo {author} {\bibfnamefont {K.}~\bibnamefont {Pierloot}},\ and\ \bibinfo {author} {\bibfnamefont {M.~F.~A.}\ \bibnamefont {Hendrickx}},\ }\bibfield  {title} {\bibinfo {title} {Embedding {Fragment} ab {Initio} {Model} {Potentials} in {CASSCF}/{CASPT2} {Calculations} of {Doped} {Solids}: {Implementation} and {Applications}},\ }\href {https://doi.org/10.1021/ct7003148} {\bibfield  {journal} {\bibinfo  {journal} {J. Chem. Theory Comp.}\ }\textbf {\bibinfo {volume} {4}},\ \bibinfo {pages} {586} (\bibinfo {year} {2008})}\BibitemShut {NoStop}%
\bibitem [{\citenamefont {Larsson}\ \emph {et~al.}(2022)\citenamefont {Larsson}, \citenamefont {Krośnicki},\ and\ \citenamefont {Veryazov}}]{Larsson2022}%
  \BibitemOpen
  \bibfield  {author} {\bibinfo {author} {\bibfnamefont {E.~D.}\ \bibnamefont {Larsson}}, \bibinfo {author} {\bibfnamefont {M.}~\bibnamefont {Krośnicki}},\ and\ \bibinfo {author} {\bibfnamefont {V.}~\bibnamefont {Veryazov}},\ }\bibfield  {title} {\bibinfo {title} {A program system for self-consistent embedded potentials for ionic crystals},\ }\href@noop {} {\bibfield  {journal} {\bibinfo  {journal} {Chem. Phys.}\ }\textbf {\bibinfo {volume} {562}},\ \bibinfo {pages} {111549} (\bibinfo {year} {2022})}\BibitemShut {NoStop}%
\bibitem [{sup()}]{suppl}%
  \BibitemOpen
  \bibinfo {note} {See Supplemental Material that includes references \onlinecite{Erba2022, usvyat2013, Pulay1980, Lee1988, Grimme2010, VilelaOliveira2019, weigend02b, 10.1063/1.455556, Celani2000, doi:10.1063/1.4940398, Handy1989} at [URL]}\BibitemShut {NoStop}%
\bibitem [{\citenamefont {Knowles}\ and\ \citenamefont {Handy}(1989)}]{Knowles89}%
  \BibitemOpen
  \bibfield  {author} {\bibinfo {author} {\bibfnamefont {P.~J.}\ \bibnamefont {Knowles}}\ and\ \bibinfo {author} {\bibfnamefont {N.~C.}\ \bibnamefont {Handy}},\ }\bibfield  {title} {\bibinfo {title} {A determinant based full configuration interaction program.},\ }\href@noop {} {\bibfield  {journal} {\bibinfo  {journal} {Comput. Phys. Commun.}\ }\textbf {\bibinfo {volume} {54}},\ \bibinfo {pages} {75} (\bibinfo {year} {1989})}\BibitemShut {NoStop}%
\bibitem [{\citenamefont {Werner}\ \emph {et~al.}(2020)\citenamefont {Werner}, \citenamefont {Knowles}, \citenamefont {Manby}, \citenamefont {Black}, \citenamefont {Doll}, \citenamefont {Heßelmann}, \citenamefont {Kats}, \citenamefont {Köhn}, \citenamefont {Korona}, \citenamefont {Kreplin}, \citenamefont {Ma}, \citenamefont {Miller}, \citenamefont {Mitrushchenkov}, \citenamefont {Peterson}, \citenamefont {Polyak}, \citenamefont {Rauhut},\ and\ \citenamefont {Sibaev}}]{molpro}%
  \BibitemOpen
  \bibfield  {author} {\bibinfo {author} {\bibfnamefont {H.-J.}\ \bibnamefont {Werner}}, \bibinfo {author} {\bibfnamefont {P.~J.}\ \bibnamefont {Knowles}}, \bibinfo {author} {\bibfnamefont {F.~R.}\ \bibnamefont {Manby}}, \bibinfo {author} {\bibfnamefont {J.~A.}\ \bibnamefont {Black}}, \bibinfo {author} {\bibfnamefont {K.}~\bibnamefont {Doll}}, \bibinfo {author} {\bibfnamefont {A.}~\bibnamefont {Heßelmann}}, \bibinfo {author} {\bibfnamefont {D.}~\bibnamefont {Kats}}, \bibinfo {author} {\bibfnamefont {A.}~\bibnamefont {Köhn}}, \bibinfo {author} {\bibfnamefont {T.}~\bibnamefont {Korona}}, \bibinfo {author} {\bibfnamefont {D.~A.}\ \bibnamefont {Kreplin}}, \bibinfo {author} {\bibfnamefont {Q.}~\bibnamefont {Ma}}, \bibinfo {author} {\bibfnamefont {T.~F.}\ \bibnamefont {Miller}}, \bibinfo {author} {\bibfnamefont {A.}~\bibnamefont {Mitrushchenkov}}, \bibinfo {author} {\bibfnamefont {K.~A.}\ \bibnamefont {Peterson}}, \bibinfo {author} {\bibfnamefont {I.}~\bibnamefont {Polyak}}, \bibinfo {author} {\bibfnamefont
  {G.}~\bibnamefont {Rauhut}},\ and\ \bibinfo {author} {\bibfnamefont {M.}~\bibnamefont {Sibaev}},\ }\bibfield  {title} {\bibinfo {title} {The molpro quantum chemistry package},\ }\href@noop {} {\bibfield  {journal} {\bibinfo  {journal} {J. Chem. Phys.}\ }\textbf {\bibinfo {volume} {152}},\ \bibinfo {pages} {144107} (\bibinfo {year} {2020})}\BibitemShut {NoStop}%
\bibitem [{\citenamefont {Kats}\ and\ \citenamefont {Manby}(2013)}]{Kats2013}%
  \BibitemOpen
  \bibfield  {author} {\bibinfo {author} {\bibfnamefont {D.}~\bibnamefont {Kats}}\ and\ \bibinfo {author} {\bibfnamefont {F.~R.}\ \bibnamefont {Manby}},\ }\bibfield  {title} {\bibinfo {title} {Communication: The distinguishable cluster approximation},\ }\href@noop {} {\bibfield  {journal} {\bibinfo  {journal} {J. Chem. Phys.}\ }\textbf {\bibinfo {volume} {139}},\ \bibinfo {pages} {021102} (\bibinfo {year} {2013})}\BibitemShut {NoStop}%
\bibitem [{\citenamefont {Kats}(2014)}]{Kats2014}%
  \BibitemOpen
  \bibfield  {author} {\bibinfo {author} {\bibfnamefont {D.}~\bibnamefont {Kats}},\ }\bibfield  {title} {\bibinfo {title} {Communication: The distinguishable cluster approximation. ii. the role of orbital relaxation},\ }\href@noop {} {\bibfield  {journal} {\bibinfo  {journal} {J. Chem. Phys.}\ }\textbf {\bibinfo {volume} {141}},\ \bibinfo {pages} {061101} (\bibinfo {year} {2014})}\BibitemShut {NoStop}%
\bibitem [{\citenamefont {Cohen}\ \emph {et~al.}(2019)\citenamefont {Cohen}, \citenamefont {Luo}, \citenamefont {Guther}, \citenamefont {Dobrautz}, \citenamefont {Tew},\ and\ \citenamefont {Alavi}}]{Cohen2019}%
  \BibitemOpen
  \bibfield  {author} {\bibinfo {author} {\bibfnamefont {A.~J.}\ \bibnamefont {Cohen}}, \bibinfo {author} {\bibfnamefont {H.}~\bibnamefont {Luo}}, \bibinfo {author} {\bibfnamefont {K.}~\bibnamefont {Guther}}, \bibinfo {author} {\bibfnamefont {W.}~\bibnamefont {Dobrautz}}, \bibinfo {author} {\bibfnamefont {D.~P.}\ \bibnamefont {Tew}},\ and\ \bibinfo {author} {\bibfnamefont {A.}~\bibnamefont {Alavi}},\ }\bibfield  {title} {\bibinfo {title} {Similarity transformation of the electronic schr\"{o}dinger equation via jastrow factorization},\ }\href@noop {} {\bibfield  {journal} {\bibinfo  {journal} {J. Chem. Phys.}\ }\textbf {\bibinfo {volume} {151}},\ \bibinfo {pages} {061101} (\bibinfo {year} {2019})}\BibitemShut {NoStop}%
\bibitem [{\citenamefont {Christlmaier}\ \emph {et~al.}(2023)\citenamefont {Christlmaier}, \citenamefont {Schraivogel}, \citenamefont {López~Ríos}, \citenamefont {Alavi},\ and\ \citenamefont {Kats}}]{Christlmaier2023}%
  \BibitemOpen
  \bibfield  {author} {\bibinfo {author} {\bibfnamefont {E.~M.~C.}\ \bibnamefont {Christlmaier}}, \bibinfo {author} {\bibfnamefont {T.}~\bibnamefont {Schraivogel}}, \bibinfo {author} {\bibfnamefont {P.}~\bibnamefont {López~Ríos}}, \bibinfo {author} {\bibfnamefont {A.}~\bibnamefont {Alavi}},\ and\ \bibinfo {author} {\bibfnamefont {D.}~\bibnamefont {Kats}},\ }\bibfield  {title} {\bibinfo {title} {xtc: An efficient treatment of three-body interactions in transcorrelated methods},\ }\href@noop {} {\bibfield  {journal} {\bibinfo  {journal} {J. Chem. Phys.}\ }\textbf {\bibinfo {volume} {159}},\ \bibinfo {pages} {014113} (\bibinfo {year} {2023})}\BibitemShut {NoStop}%
\bibitem [{\citenamefont {Erba}\ \emph {et~al.}(2022)\citenamefont {Erba}, \citenamefont {Desmarais}, \citenamefont {Casassa}, \citenamefont {Civalleri}, \citenamefont {Donà}, \citenamefont {Bush}, \citenamefont {Searle}, \citenamefont {Maschio}, \citenamefont {Edith-Daga}, \citenamefont {Cossard}, \citenamefont {Ribaldone}, \citenamefont {Ascrizzi}, \citenamefont {Marana}, \citenamefont {Flament},\ and\ \citenamefont {Kirtman}}]{Erba2022}%
  \BibitemOpen
  \bibfield  {author} {\bibinfo {author} {\bibfnamefont {A.}~\bibnamefont {Erba}}, \bibinfo {author} {\bibfnamefont {J.~K.}\ \bibnamefont {Desmarais}}, \bibinfo {author} {\bibfnamefont {S.}~\bibnamefont {Casassa}}, \bibinfo {author} {\bibfnamefont {B.}~\bibnamefont {Civalleri}}, \bibinfo {author} {\bibfnamefont {L.}~\bibnamefont {Donà}}, \bibinfo {author} {\bibfnamefont {I.~J.}\ \bibnamefont {Bush}}, \bibinfo {author} {\bibfnamefont {B.}~\bibnamefont {Searle}}, \bibinfo {author} {\bibfnamefont {L.}~\bibnamefont {Maschio}}, \bibinfo {author} {\bibfnamefont {L.}~\bibnamefont {Edith-Daga}}, \bibinfo {author} {\bibfnamefont {A.}~\bibnamefont {Cossard}}, \bibinfo {author} {\bibfnamefont {C.}~\bibnamefont {Ribaldone}}, \bibinfo {author} {\bibfnamefont {E.}~\bibnamefont {Ascrizzi}}, \bibinfo {author} {\bibfnamefont {N.~L.}\ \bibnamefont {Marana}}, \bibinfo {author} {\bibfnamefont {J.-P.}\ \bibnamefont {Flament}},\ and\ \bibinfo {author} {\bibfnamefont {B.}~\bibnamefont {Kirtman}},\ }\bibfield  {title} {\bibinfo
  {title} {Crystal23: A program for computational solid state physics and chemistry},\ }\href@noop {} {\bibfield  {journal} {\bibinfo  {journal} {J. Chem. Theory Comput.}\ }\textbf {\bibinfo {volume} {19}},\ \bibinfo {pages} {6891} (\bibinfo {year} {2022})}\BibitemShut {NoStop}%
\bibitem [{\citenamefont {Usvyat}(2013)}]{usvyat2013}%
  \BibitemOpen
  \bibfield  {author} {\bibinfo {author} {\bibfnamefont {D.}~\bibnamefont {Usvyat}},\ }\bibfield  {title} {\bibinfo {title} {Linear-scaling explicitly correlated treatment of solids: Periodic local mp2-f12 method},\ }\href@noop {} {\bibfield  {journal} {\bibinfo  {journal} {J. Chem. Phys.}\ }\textbf {\bibinfo {volume} {139}},\ \bibinfo {pages} {194101} (\bibinfo {year} {2013})}\BibitemShut {NoStop}%
\bibitem [{\citenamefont {Pulay}(1980)}]{Pulay1980}%
  \BibitemOpen
  \bibfield  {author} {\bibinfo {author} {\bibfnamefont {P.}~\bibnamefont {Pulay}},\ }\bibfield  {title} {\bibinfo {title} {Convergence acceleration of iterative sequences. the case of scf iteration},\ }\href@noop {} {\bibfield  {journal} {\bibinfo  {journal} {Chem. Phys. Lett.}\ }\textbf {\bibinfo {volume} {73}},\ \bibinfo {pages} {393} (\bibinfo {year} {1980})}\BibitemShut {NoStop}%
\bibitem [{\citenamefont {Lee}\ \emph {et~al.}(1988)\citenamefont {Lee}, \citenamefont {Yang},\ and\ \citenamefont {Parr}}]{Lee1988}%
  \BibitemOpen
  \bibfield  {author} {\bibinfo {author} {\bibfnamefont {C.}~\bibnamefont {Lee}}, \bibinfo {author} {\bibfnamefont {W.}~\bibnamefont {Yang}},\ and\ \bibinfo {author} {\bibfnamefont {R.~G.}\ \bibnamefont {Parr}},\ }\bibfield  {title} {\bibinfo {title} {Development of the colle-salvetti correlation-energy formula into a functional of the electron density},\ }\href@noop {} {\bibfield  {journal} {\bibinfo  {journal} {Phys. Rev. B}\ }\textbf {\bibinfo {volume} {37}},\ \bibinfo {pages} {785} (\bibinfo {year} {1988})}\BibitemShut {NoStop}%
\bibitem [{\citenamefont {Grimme}\ \emph {et~al.}(2010)\citenamefont {Grimme}, \citenamefont {Antony}, \citenamefont {Ehrlich},\ and\ \citenamefont {Krieg}}]{Grimme2010}%
  \BibitemOpen
  \bibfield  {author} {\bibinfo {author} {\bibfnamefont {S.}~\bibnamefont {Grimme}}, \bibinfo {author} {\bibfnamefont {J.}~\bibnamefont {Antony}}, \bibinfo {author} {\bibfnamefont {S.}~\bibnamefont {Ehrlich}},\ and\ \bibinfo {author} {\bibfnamefont {H.}~\bibnamefont {Krieg}},\ }\bibfield  {title} {\bibinfo {title} {A consistent and accurateab initioparametrization of density functional dispersion correction (dft-d) for the 94 elements h-pu},\ }\href@noop {} {\bibfield  {journal} {\bibinfo  {journal} {J. Chem. Phys.}\ }\textbf {\bibinfo {volume} {132}},\ \bibinfo {pages} {154104} (\bibinfo {year} {2010})}\BibitemShut {NoStop}%
\bibitem [{\citenamefont {Vilela~Oliveira}\ \emph {et~al.}(2019)\citenamefont {Vilela~Oliveira}, \citenamefont {Laun}, \citenamefont {Peintinger},\ and\ \citenamefont {Bredow}}]{VilelaOliveira2019}%
  \BibitemOpen
  \bibfield  {author} {\bibinfo {author} {\bibfnamefont {D.}~\bibnamefont {Vilela~Oliveira}}, \bibinfo {author} {\bibfnamefont {J.}~\bibnamefont {Laun}}, \bibinfo {author} {\bibfnamefont {M.~F.}\ \bibnamefont {Peintinger}},\ and\ \bibinfo {author} {\bibfnamefont {T.}~\bibnamefont {Bredow}},\ }\bibfield  {title} {\bibinfo {title} {Bsse‐correction scheme for consistent gaussian basis sets of double‐ and triple‐zeta valence with polarization quality for solid‐state calculations},\ }\href@noop {} {\bibfield  {journal} {\bibinfo  {journal} {J. Comput. Chem.}\ }\textbf {\bibinfo {volume} {40}},\ \bibinfo {pages} {2364} (\bibinfo {year} {2019})}\BibitemShut {NoStop}%
\bibitem [{\citenamefont {Weigend}\ \emph {et~al.}(2002)\citenamefont {Weigend}, \citenamefont {K\"ohn},\ and\ \citenamefont {H\"attig}}]{weigend02b}%
  \BibitemOpen
  \bibfield  {author} {\bibinfo {author} {\bibfnamefont {F.}~\bibnamefont {Weigend}}, \bibinfo {author} {\bibfnamefont {A.}~\bibnamefont {K\"ohn}},\ and\ \bibinfo {author} {\bibfnamefont {C.}~\bibnamefont {H\"attig}},\ }\bibfield  {title} {\bibinfo {title} {Efficient use of the correlation consistent basis sets in resolution of the identity mp2 calculations},\ }\href@noop {} {\bibfield  {journal} {\bibinfo  {journal} {J. Chem. Phys.}\ }\textbf {\bibinfo {volume} {116}},\ \bibinfo {pages} {3175} (\bibinfo {year} {2002})}\BibitemShut {NoStop}%
\bibitem [{\citenamefont {Werner}\ and\ \citenamefont {Knowles}(1988)}]{10.1063/1.455556}%
  \BibitemOpen
  \bibfield  {author} {\bibinfo {author} {\bibfnamefont {H.}~\bibnamefont {Werner}}\ and\ \bibinfo {author} {\bibfnamefont {P.~J.}\ \bibnamefont {Knowles}},\ }\bibfield  {title} {\bibinfo {title} {{An efficient internally contracted multiconfiguration–reference configuration interaction method}},\ }\href@noop {} {\bibfield  {journal} {\bibinfo  {journal} {J. Chem. Phys.}\ }\textbf {\bibinfo {volume} {89}},\ \bibinfo {pages} {5803} (\bibinfo {year} {1988})}\BibitemShut {NoStop}%
\bibitem [{\citenamefont {Celani}\ and\ \citenamefont {Werner}(2000)}]{Celani2000}%
  \BibitemOpen
  \bibfield  {author} {\bibinfo {author} {\bibfnamefont {P.}~\bibnamefont {Celani}}\ and\ \bibinfo {author} {\bibfnamefont {H.-J.}\ \bibnamefont {Werner}},\ }\bibfield  {title} {\bibinfo {title} {Multireference perturbation theory for large restricted and selected active space reference wave functions},\ }\href@noop {} {\bibfield  {journal} {\bibinfo  {journal} {J. Chem. Phys.}\ }\textbf {\bibinfo {volume} {112}},\ \bibinfo {pages} {5546} (\bibinfo {year} {2000})}\BibitemShut {NoStop}%
\bibitem [{\citenamefont {Kats}(2016)}]{doi:10.1063/1.4940398}%
  \BibitemOpen
  \bibfield  {author} {\bibinfo {author} {\bibfnamefont {D.}~\bibnamefont {Kats}},\ }\bibfield  {title} {\bibinfo {title} {The distinguishable cluster approach from a screened coulomb formalism},\ }\bibfield  {journal} {\bibinfo  {journal} {J. Chem. Phys.}\ }\textbf {\bibinfo {volume} {144}},\ \href {https://doi.org/10.1063/1.4940398} {10.1063/1.4940398} (\bibinfo {year} {2016})\BibitemShut {NoStop}%
\bibitem [{\citenamefont {Handy}\ \emph {et~al.}(1989)\citenamefont {Handy}, \citenamefont {Pople}, \citenamefont {Head-Gordon}, \citenamefont {Raghavachari},\ and\ \citenamefont {Trucks}}]{Handy1989}%
  \BibitemOpen
  \bibfield  {author} {\bibinfo {author} {\bibfnamefont {N.~C.}\ \bibnamefont {Handy}}, \bibinfo {author} {\bibfnamefont {J.~A.}\ \bibnamefont {Pople}}, \bibinfo {author} {\bibfnamefont {M.}~\bibnamefont {Head-Gordon}}, \bibinfo {author} {\bibfnamefont {K.}~\bibnamefont {Raghavachari}},\ and\ \bibinfo {author} {\bibfnamefont {G.~W.}\ \bibnamefont {Trucks}},\ }\bibfield  {title} {\bibinfo {title} {Size-consistent brueckner theory limited to double substitutions},\ }\href@noop {} {\bibfield  {journal} {\bibinfo  {journal} {Chem. Phys. Lett.}\ }\textbf {\bibinfo {volume} {164}},\ \bibinfo {pages} {185} (\bibinfo {year} {1989})}\BibitemShut {NoStop}%
\end{thebibliography}%


\begin{thebibliography}{17}%
\makeatletter
\providecommand \@ifxundefined [1]{%
 \@ifx{#1\undefined}
}%
\providecommand \@ifnum [1]{%
 \ifnum #1\expandafter \@firstoftwo
 \else \expandafter \@secondoftwo
 \fi
}%
\providecommand \@ifx [1]{%
 \ifx #1\expandafter \@firstoftwo
 \else \expandafter \@secondoftwo
 \fi
}%
\providecommand \natexlab [1]{#1}%
\providecommand \enquote  [1]{``#1''}%
\providecommand \bibnamefont  [1]{#1}%
\providecommand \bibfnamefont [1]{#1}%
\providecommand \citenamefont [1]{#1}%
\providecommand \href@noop [0]{\@secondoftwo}%
\providecommand \href [0]{\begingroup \@sanitize@url \@href}%
\providecommand \@href[1]{\@@startlink{#1}\@@href}%
\providecommand \@@href[1]{\endgroup#1\@@endlink}%
\providecommand \@sanitize@url [0]{\catcode `\\12\catcode `\$12\catcode `\&12\catcode `\#12\catcode `\^12\catcode `\_12\catcode `\%12\relax}%
\providecommand \@@startlink[1]{}%
\providecommand \@@endlink[0]{}%
\providecommand \url  [0]{\begingroup\@sanitize@url \@url }%
\providecommand \@url [1]{\endgroup\@href {#1}{\urlprefix }}%
\providecommand \urlprefix  [0]{URL }%
\providecommand \Eprint [0]{\href }%
\providecommand \doibase [0]{https://doi.org/}%
\providecommand \selectlanguage [0]{\@gobble}%
\providecommand \bibinfo  [0]{\@secondoftwo}%
\providecommand \bibfield  [0]{\@secondoftwo}%
\providecommand \translation [1]{[#1]}%
\providecommand \BibitemOpen [0]{}%
\providecommand \bibitemStop [0]{}%
\providecommand \bibitemNoStop [0]{.\EOS\space}%
\providecommand \EOS [0]{\spacefactor3000\relax}%
\providecommand \BibitemShut  [1]{\csname bibitem#1\endcsname}%
\let\auto@bib@innerbib\@empty
\bibitem [{\citenamefont {Erba}\ \emph {et~al.}(2022)\citenamefont {Erba}, \citenamefont {Desmarais}, \citenamefont {Casassa}, \citenamefont {Civalleri}, \citenamefont {Donà}, \citenamefont {Bush}, \citenamefont {Searle}, \citenamefont {Maschio}, \citenamefont {Edith-Daga}, \citenamefont {Cossard}, \citenamefont {Ribaldone}, \citenamefont {Ascrizzi}, \citenamefont {Marana}, \citenamefont {Flament},\ and\ \citenamefont {Kirtman}}]{Erba2022}%
  \BibitemOpen
  \bibfield  {author} {\bibinfo {author} {\bibfnamefont {A.}~\bibnamefont {Erba}}, \bibinfo {author} {\bibfnamefont {J.~K.}\ \bibnamefont {Desmarais}}, \bibinfo {author} {\bibfnamefont {S.}~\bibnamefont {Casassa}}, \bibinfo {author} {\bibfnamefont {B.}~\bibnamefont {Civalleri}}, \bibinfo {author} {\bibfnamefont {L.}~\bibnamefont {Donà}}, \bibinfo {author} {\bibfnamefont {I.~J.}\ \bibnamefont {Bush}}, \bibinfo {author} {\bibfnamefont {B.}~\bibnamefont {Searle}}, \bibinfo {author} {\bibfnamefont {L.}~\bibnamefont {Maschio}}, \bibinfo {author} {\bibfnamefont {L.}~\bibnamefont {Edith-Daga}}, \bibinfo {author} {\bibfnamefont {A.}~\bibnamefont {Cossard}}, \bibinfo {author} {\bibfnamefont {C.}~\bibnamefont {Ribaldone}}, \bibinfo {author} {\bibfnamefont {E.}~\bibnamefont {Ascrizzi}}, \bibinfo {author} {\bibfnamefont {N.~L.}\ \bibnamefont {Marana}}, \bibinfo {author} {\bibfnamefont {J.-P.}\ \bibnamefont {Flament}},\ and\ \bibinfo {author} {\bibfnamefont {B.}~\bibnamefont {Kirtman}},\ }\bibfield  {title} {\bibinfo
  {title} {Crystal23: A program for computational solid state physics and chemistry},\ }\href@noop {} {\bibfield  {journal} {\bibinfo  {journal} {J. Chem. Theory Comput.}\ }\textbf {\bibinfo {volume} {19}},\ \bibinfo {pages} {6891} (\bibinfo {year} {2022})}\BibitemShut {NoStop}%
\bibitem [{\citenamefont {Usvyat}\ \emph {et~al.}(2018)\citenamefont {Usvyat}, \citenamefont {Maschio},\ and\ \citenamefont {Sch{\"u}tz}}]{usvyat18}%
  \BibitemOpen
  \bibfield  {author} {\bibinfo {author} {\bibfnamefont {D.}~\bibnamefont {Usvyat}}, \bibinfo {author} {\bibfnamefont {L.}~\bibnamefont {Maschio}},\ and\ \bibinfo {author} {\bibfnamefont {M.}~\bibnamefont {Sch{\"u}tz}},\ }\bibfield  {title} {\bibinfo {title} {Periodic and fragment models based on the local correlation approach},\ }\href@noop {} {\bibfield  {journal} {\bibinfo  {journal} {WIREs: Comput. Mol. Sci.}\ }\textbf {\bibinfo {volume} {8}},\ \bibinfo {pages} {e1357} (\bibinfo {year} {2018})}\BibitemShut {NoStop}%
\bibitem [{\citenamefont {Pisani}\ \emph {et~al.}(2012)\citenamefont {Pisani}, \citenamefont {Schütz}, \citenamefont {Casassa}, \citenamefont {Usvyat}, \citenamefont {Maschio}, \citenamefont {Lorenz},\ and\ \citenamefont {Erba}}]{Pisani2012}%
  \BibitemOpen
  \bibfield  {author} {\bibinfo {author} {\bibfnamefont {C.}~\bibnamefont {Pisani}}, \bibinfo {author} {\bibfnamefont {M.}~\bibnamefont {Schütz}}, \bibinfo {author} {\bibfnamefont {S.}~\bibnamefont {Casassa}}, \bibinfo {author} {\bibfnamefont {D.}~\bibnamefont {Usvyat}}, \bibinfo {author} {\bibfnamefont {L.}~\bibnamefont {Maschio}}, \bibinfo {author} {\bibfnamefont {M.}~\bibnamefont {Lorenz}},\ and\ \bibinfo {author} {\bibfnamefont {A.}~\bibnamefont {Erba}},\ }\bibfield  {title} {\bibinfo {title} {Cryscor: A program for the post-hartree-fock treatment of periodic systems},\ }\href@noop {} {\bibfield  {journal} {\bibinfo  {journal} {Phys. Chem. Chem. Phys.}\ }\textbf {\bibinfo {volume} {14}},\ \bibinfo {pages} {7615} (\bibinfo {year} {2012})}\BibitemShut {NoStop}%
\bibitem [{\citenamefont {Usvyat}(2013)}]{usvyat2013}%
  \BibitemOpen
  \bibfield  {author} {\bibinfo {author} {\bibfnamefont {D.}~\bibnamefont {Usvyat}},\ }\bibfield  {title} {\bibinfo {title} {Linear-scaling explicitly correlated treatment of solids: Periodic local mp2-f12 method},\ }\href@noop {} {\bibfield  {journal} {\bibinfo  {journal} {J. Chem. Phys.}\ }\textbf {\bibinfo {volume} {139}},\ \bibinfo {pages} {194101} (\bibinfo {year} {2013})}\BibitemShut {NoStop}%
\bibitem [{\citenamefont {Christlmaier}\ \emph {et~al.}(2022)\citenamefont {Christlmaier}, \citenamefont {Kats}, \citenamefont {Alavi},\ and\ \citenamefont {Usvyat}}]{christlmaier21}%
  \BibitemOpen
  \bibfield  {author} {\bibinfo {author} {\bibfnamefont {E.~M.}\ \bibnamefont {Christlmaier}}, \bibinfo {author} {\bibfnamefont {D.}~\bibnamefont {Kats}}, \bibinfo {author} {\bibfnamefont {A.}~\bibnamefont {Alavi}},\ and\ \bibinfo {author} {\bibfnamefont {D.}~\bibnamefont {Usvyat}},\ }\bibfield  {title} {\bibinfo {title} {Full configuration interaction quantum monte carlo treatment of fragments embedded in a periodic mean field},\ }\href@noop {} {\bibfield  {journal} {\bibinfo  {journal} {J. Chem. Phys.}\ }\textbf {\bibinfo {volume} {156}},\ \bibinfo {pages} {074109} (\bibinfo {year} {2022})}\BibitemShut {NoStop}%
\bibitem [{\citenamefont {Pulay}(1980)}]{Pulay1980}%
  \BibitemOpen
  \bibfield  {author} {\bibinfo {author} {\bibfnamefont {P.}~\bibnamefont {Pulay}},\ }\bibfield  {title} {\bibinfo {title} {Convergence acceleration of iterative sequences. the case of scf iteration},\ }\href@noop {} {\bibfield  {journal} {\bibinfo  {journal} {Chem. Phys. Lett.}\ }\textbf {\bibinfo {volume} {73}},\ \bibinfo {pages} {393} (\bibinfo {year} {1980})}\BibitemShut {NoStop}%
\bibitem [{\citenamefont {Knowles}\ and\ \citenamefont {Handy}(1989)}]{Knowles89}%
  \BibitemOpen
  \bibfield  {author} {\bibinfo {author} {\bibfnamefont {P.~J.}\ \bibnamefont {Knowles}}\ and\ \bibinfo {author} {\bibfnamefont {N.~C.}\ \bibnamefont {Handy}},\ }\bibfield  {title} {\bibinfo {title} {A determinant based full configuration interaction program.},\ }\href@noop {} {\bibfield  {journal} {\bibinfo  {journal} {Comput. Phys. Commun.}\ }\textbf {\bibinfo {volume} {54}},\ \bibinfo {pages} {75} (\bibinfo {year} {1989})}\BibitemShut {NoStop}%
\bibitem [{\citenamefont {Lee}\ \emph {et~al.}(1988)\citenamefont {Lee}, \citenamefont {Yang},\ and\ \citenamefont {Parr}}]{Lee1988}%
  \BibitemOpen
  \bibfield  {author} {\bibinfo {author} {\bibfnamefont {C.}~\bibnamefont {Lee}}, \bibinfo {author} {\bibfnamefont {W.}~\bibnamefont {Yang}},\ and\ \bibinfo {author} {\bibfnamefont {R.~G.}\ \bibnamefont {Parr}},\ }\bibfield  {title} {\bibinfo {title} {Development of the colle-salvetti correlation-energy formula into a functional of the electron density},\ }\href@noop {} {\bibfield  {journal} {\bibinfo  {journal} {Phys. Rev. B}\ }\textbf {\bibinfo {volume} {37}},\ \bibinfo {pages} {785} (\bibinfo {year} {1988})}\BibitemShut {NoStop}%
\bibitem [{\citenamefont {Grimme}\ \emph {et~al.}(2010)\citenamefont {Grimme}, \citenamefont {Antony}, \citenamefont {Ehrlich},\ and\ \citenamefont {Krieg}}]{Grimme2010}%
  \BibitemOpen
  \bibfield  {author} {\bibinfo {author} {\bibfnamefont {S.}~\bibnamefont {Grimme}}, \bibinfo {author} {\bibfnamefont {J.}~\bibnamefont {Antony}}, \bibinfo {author} {\bibfnamefont {S.}~\bibnamefont {Ehrlich}},\ and\ \bibinfo {author} {\bibfnamefont {H.}~\bibnamefont {Krieg}},\ }\bibfield  {title} {\bibinfo {title} {A consistent and accurateab initioparametrization of density functional dispersion correction (dft-d) for the 94 elements h-pu},\ }\href@noop {} {\bibfield  {journal} {\bibinfo  {journal} {J. Chem. Phys.}\ }\textbf {\bibinfo {volume} {132}},\ \bibinfo {pages} {154104} (\bibinfo {year} {2010})}\BibitemShut {NoStop}%
\bibitem [{\citenamefont {Vilela~Oliveira}\ \emph {et~al.}(2019)\citenamefont {Vilela~Oliveira}, \citenamefont {Laun}, \citenamefont {Peintinger},\ and\ \citenamefont {Bredow}}]{VilelaOliveira2019}%
  \BibitemOpen
  \bibfield  {author} {\bibinfo {author} {\bibfnamefont {D.}~\bibnamefont {Vilela~Oliveira}}, \bibinfo {author} {\bibfnamefont {J.}~\bibnamefont {Laun}}, \bibinfo {author} {\bibfnamefont {M.~F.}\ \bibnamefont {Peintinger}},\ and\ \bibinfo {author} {\bibfnamefont {T.}~\bibnamefont {Bredow}},\ }\bibfield  {title} {\bibinfo {title} {Bsse‐correction scheme for consistent gaussian basis sets of double‐ and triple‐zeta valence with polarization quality for solid‐state calculations},\ }\href@noop {} {\bibfield  {journal} {\bibinfo  {journal} {J. Comput. Chem.}\ }\textbf {\bibinfo {volume} {40}},\ \bibinfo {pages} {2364} (\bibinfo {year} {2019})}\BibitemShut {NoStop}%
\bibitem [{\citenamefont {Weigend}\ \emph {et~al.}(2002)\citenamefont {Weigend}, \citenamefont {K\"ohn},\ and\ \citenamefont {H\"attig}}]{weigend02b}%
  \BibitemOpen
  \bibfield  {author} {\bibinfo {author} {\bibfnamefont {F.}~\bibnamefont {Weigend}}, \bibinfo {author} {\bibfnamefont {A.}~\bibnamefont {K\"ohn}},\ and\ \bibinfo {author} {\bibfnamefont {C.}~\bibnamefont {H\"attig}},\ }\bibfield  {title} {\bibinfo {title} {Efficient use of the correlation consistent basis sets in resolution of the identity mp2 calculations},\ }\href@noop {} {\bibfield  {journal} {\bibinfo  {journal} {J. Chem. Phys.}\ }\textbf {\bibinfo {volume} {116}},\ \bibinfo {pages} {3175} (\bibinfo {year} {2002})}\BibitemShut {NoStop}%
\bibitem [{\citenamefont {Werner}\ and\ \citenamefont {Knowles}(1988)}]{10.1063/1.455556}%
  \BibitemOpen
  \bibfield  {author} {\bibinfo {author} {\bibfnamefont {H.}~\bibnamefont {Werner}}\ and\ \bibinfo {author} {\bibfnamefont {P.~J.}\ \bibnamefont {Knowles}},\ }\bibfield  {title} {\bibinfo {title} {{An efficient internally contracted multiconfiguration–reference configuration interaction method}},\ }\href@noop {} {\bibfield  {journal} {\bibinfo  {journal} {J. Chem. Phys.}\ }\textbf {\bibinfo {volume} {89}},\ \bibinfo {pages} {5803} (\bibinfo {year} {1988})}\BibitemShut {NoStop}%
\bibitem [{\citenamefont {Celani}\ and\ \citenamefont {Werner}(2000)}]{Celani2000}%
  \BibitemOpen
  \bibfield  {author} {\bibinfo {author} {\bibfnamefont {P.}~\bibnamefont {Celani}}\ and\ \bibinfo {author} {\bibfnamefont {H.-J.}\ \bibnamefont {Werner}},\ }\bibfield  {title} {\bibinfo {title} {Multireference perturbation theory for large restricted and selected active space reference wave functions},\ }\href@noop {} {\bibfield  {journal} {\bibinfo  {journal} {J. Chem. Phys.}\ }\textbf {\bibinfo {volume} {112}},\ \bibinfo {pages} {5546} (\bibinfo {year} {2000})}\BibitemShut {NoStop}%
\bibitem [{\citenamefont {Kats}\ and\ \citenamefont {Manby}(2013)}]{Kats2013}%
  \BibitemOpen
  \bibfield  {author} {\bibinfo {author} {\bibfnamefont {D.}~\bibnamefont {Kats}}\ and\ \bibinfo {author} {\bibfnamefont {F.~R.}\ \bibnamefont {Manby}},\ }\bibfield  {title} {\bibinfo {title} {Communication: The distinguishable cluster approximation},\ }\href@noop {} {\bibfield  {journal} {\bibinfo  {journal} {J. Chem. Phys.}\ }\textbf {\bibinfo {volume} {139}},\ \bibinfo {pages} {021102} (\bibinfo {year} {2013})}\BibitemShut {NoStop}%
\bibitem [{\citenamefont {Kats}(2016)}]{doi:10.1063/1.4940398}%
  \BibitemOpen
  \bibfield  {author} {\bibinfo {author} {\bibfnamefont {D.}~\bibnamefont {Kats}},\ }\bibfield  {title} {\bibinfo {title} {The distinguishable cluster approach from a screened coulomb formalism},\ }\bibfield  {journal} {\bibinfo  {journal} {J. Chem. Phys.}\ }\textbf {\bibinfo {volume} {144}},\ \href {https://doi.org/10.1063/1.4940398} {10.1063/1.4940398} (\bibinfo {year} {2016})\BibitemShut {NoStop}%
\bibitem [{\citenamefont {Kats}(2014)}]{Kats2014}%
  \BibitemOpen
  \bibfield  {author} {\bibinfo {author} {\bibfnamefont {D.}~\bibnamefont {Kats}},\ }\bibfield  {title} {\bibinfo {title} {Communication: The distinguishable cluster approximation. ii. the role of orbital relaxation},\ }\href@noop {} {\bibfield  {journal} {\bibinfo  {journal} {J. Chem. Phys.}\ }\textbf {\bibinfo {volume} {141}},\ \bibinfo {pages} {061101} (\bibinfo {year} {2014})}\BibitemShut {NoStop}%
\bibitem [{\citenamefont {Handy}\ \emph {et~al.}(1989)\citenamefont {Handy}, \citenamefont {Pople}, \citenamefont {Head-Gordon}, \citenamefont {Raghavachari},\ and\ \citenamefont {Trucks}}]{Handy1989}%
  \BibitemOpen
  \bibfield  {author} {\bibinfo {author} {\bibfnamefont {N.~C.}\ \bibnamefont {Handy}}, \bibinfo {author} {\bibfnamefont {J.~A.}\ \bibnamefont {Pople}}, \bibinfo {author} {\bibfnamefont {M.}~\bibnamefont {Head-Gordon}}, \bibinfo {author} {\bibfnamefont {K.}~\bibnamefont {Raghavachari}},\ and\ \bibinfo {author} {\bibfnamefont {G.~W.}\ \bibnamefont {Trucks}},\ }\bibfield  {title} {\bibinfo {title} {Size-consistent brueckner theory limited to double substitutions},\ }\href@noop {} {\bibfield  {journal} {\bibinfo  {journal} {Chem. Phys. Lett.}\ }\textbf {\bibinfo {volume} {164}},\ \bibinfo {pages} {185} (\bibinfo {year} {1989})}\BibitemShut {NoStop}%
\end{thebibliography}%

\end{document}


\title{Supplemental Material for ``Aperiodic defects in periodic solids''}

\author{Robert H. Lavroff}
\affiliation{Department of Chemistry and Biochemistry, University of California Los Angeles, Los Angeles, California, USA}
\author{Daniel Kats}
\affiliation{Max Planck Institute for Solid State Research, Heisenbergstra{\ss}e 1, D-70569 Stuttgart, Germany}
\author{Lorenzo Maschio}
\affiliation{Dipartimento di Chimica, Universit\`a di Torino, Torino, Italy}
\author{Nikolay A. Bogdanov}
\affiliation{Max Planck Institute for Solid State Research,
  Heisenbergstra{\ss}e 1, D-70569 Stuttgart, Germany}
\author{Ali Alavi}
\affiliation{Max Planck Institute for Solid State Research, Heisenbergstra{\ss}e 1, D-70569 Stuttgart, Germany}
\affiliation{Yusuf Hamied Department of Chemistry, University of Cambridge, Lensfield Road, Cambridge CB2 1EW, United Kingdom}
\author{Anastassia N. Alexandrova}
\affiliation{Department of Chemistry and Biochemistry, University of California Los Angeles, Los Angeles, California, USA}
\affiliation{Department of Materials Science and Engineering, University of California Los Angeles, Los Angeles, California, USA}
\affiliation{California Nanoscience Institute (CNSI), Los Angeles, California, USA}
\author{Denis Usvyat}
 \email{denis.usvyat@hu-berlin.de}
\affiliation{Institut f{\"u}r Chemie, Humboldt-Universit{\"a}t zu Berlin, Brook-Taylor-Str. 2, D-12489 Berlin, Germany}
\date{\today}


\maketitle

\section{HF energy of an embedded fragment}
The starting point is a model of a crystal's fragment, defined in terms of
local orbitals, electrons, and nuclei, embedded in the mean field of the rest
of the crystal. At first, we consider the case when the geometry of the fragment remains the same as in the periodic structure. This model is referred to in the main text as the frozen fragment. In our approach we define the HF energy of such a fragment as its mean-field energy as the physical object subjected to embedding. That is, the HF energy  of the isolated fragment plus the energy of the interaction between the
fragment and the environment:
\begin{widetext}
\begin{eqnarray}
E^{\rm frozen \, frag}_{\rm HF}&=&2\sum_{i\in{\rm frag}}\left< i\left| -{1\over 2} \nabla^2\right| i\right>
- 2\sum_{i\in{\rm frag}}\left< i\left| \sum_{K\in{\rm frag}}{Z_K \over |{\bf r}-{\bf
    R}_K|}\right|i\right> 
+{1 \over 2}\sum_{i\in{\rm frag}}\sum_{j\in{\rm frag}}\left[4\left(ii|jj\right)-2\left(ij|ji\right)\right]\nonumber\\
&&+{1 \over 2}\sum_{L\in{\rm frag}}\sum_{K\in{\rm frag}}\phantom{}^{'}{Z_LZ_K \over |{\bf R}_L-{\bf
    R}_K|}
- 2\sum_{i\notin{\rm frag}}\left< i\left| \sum_{K\in{\rm frag}}{Z_K \over |{\bf r}-{\bf
    R}_K|}\right|i\right>
- 2\sum_{i\in{\rm frag}}\left< i\left| \sum_{K\notin{\rm frag}}{Z_K \over |{\bf r}-{\bf
    R}_K|}\right|i\right>
\nonumber\\
&&
+\sum_{i\in{\rm frag}}\sum_{j\notin{\rm
    frag}}\left[4\left(ii|jj\right)-2\left(ij|ji\right)\right]
+\sum_{L\notin{\rm frag}}\sum_{K\in{\rm frag}}{Z_LZ_K \over |{\bf R}_L-{\bf
    R}_K|}.\label{eq:frozenfrag_E}
\end{eqnarray}
\end{widetext}
A primed summation sign denotes exclusion of the terms with two identical summation indices. The first four terms in eq. (\ref{eq:frozenfrag_E}) form the HF energy
of an isolated fragment; the 5th and 6th terms give the energy
of the Coulomb attraction between the electrons or nuclei of the environment
and nuclei or electrons of the fragment, respectively; the 7th term is the
Coulomb and exchange contributions due to the interaction between the
electrons of the environment and the fragment; and finally the 8th term is the
Coulomb repulsion between the nuclei of the environment and the fragment.


By noting that the terms 2, 4 (taken twice), 5 and 8  together can be
rewritten via the electrostatic potential of the periodic system  $V\left({\bf
  r}\right)$ at the locations of the fragment nuclei:
\begin{eqnarray}
&&Z_K\left[-2\sum_{i}\left< i\left| {1 \over |{\bf r}-{\bf
      R}_K|}\right|i\right>+\sum_L\phantom{}^{'} {Z_L \over |{\bf R}_K-{\bf R}_L|}\right]\nonumber\\
&&=Z_K\cdot V\left({\bf R}_K\right)\label{eq:electr_pot}
\end{eqnarray}
and regrouping other terms, one can simplify expression (\ref{eq:frozenfrag_E}) to:
\begin{widetext}
\begin{eqnarray}
  E^{\rm frozen\, frag}_{\rm HF}&=&2\sum_{i\in{\rm frag}}h^{\rm frozen\, frag}_{ii}
+\sum_{i\in{\rm frag}}\sum_{j\in{\rm frag}}
      \left[2\left(ii|jj\right)-\left(ij|ji\right)\right]\nonumber\\
       &&+ \sum_{K\in{\rm frag}} Z_K\cdot V\left({\bf R}_K\right) 
-{1 \over 2}\sum_{K\in{\rm frag}}\sum_{L\in{\rm frag}}\phantom{}^{'}{Z_KZ_L \over |{\bf R}_K-{\bf
   R}_L|}
   - 2\sum_{i\in{\rm frag}}\left< i\left| -\sum_{K\in{\rm frag}}{Z_K \over |{\bf r}-{\bf
    R}_K|}\right|i\right>\nonumber\\  
                                &=&
{1 \over 2}\sum_{i\in{\rm frag}}\left[2h^{\rm frozen\, frag}_{ii}+2F^{\rm
                                    frozen\, frag}_{ii}\right]+E^{\rm frozen\, frag}_{\rm nuc}.\label{eq:frozenfrag_E2}
\end{eqnarray}
\end{widetext}
The fragment Hamiltonian within the frozen fragment model is given by eq. (1) of the main text. Eq. (\ref{eq:frozenfrag_E2}) shows that the HF energy of the frozen embedded fragment can be defined
 via the fragment one-electron Hamiltonian, fragment
 Fock matrix
\begin{eqnarray}
  F^{\rm frozen\, frag}_{\mu\nu}&=&h^{\rm frozen\,frag}_{\mu\nu}\nonumber\\&&+
\sum_{i\in{\rm frag}}\sum_{i\in{\rm frag}}
  \left[2\left(\mu\nu|ii\right)-\left(\mu i|i\nu\right)\right]\label{eq:frozenfrag_F}
\end{eqnarray}
and an effective ``nuclear energy''
\begin{widetext}
\begin{eqnarray}
 E^{\rm frozen\, frag}_{\rm nuc}&=&\sum_{K\in{\rm frag}} Z_K\cdot V\left({\bf R}_K\right) 
-{1 \over 2}\sum_{K\in{\rm frag}}\sum_{L\in{\rm frag}}\phantom{}^{'}{Z_KZ_L \over |{\bf R}_K-{\bf
   R}_L|}
   - 2\sum_{i\in{\rm frag}}\left< i\left| -\sum_{K\in{\rm frag}}{Z_K \over |{\bf r}-{\bf
    R}_K|}\right|i\right>.
\end{eqnarray} 
\end{widetext}

Now we introduce a defect inside the fragment. This is done by removing, adding, and/or substituting nuclei together with the corresponding basis functions (or introducing ghost atoms) and updating the number of electrons (we re-emphasize, the fragment need not be neutral). The HF energy must be modified to include (i) the change in the position of the nuclei and (ii) to adapt to the new fragment orbitals as the new fragment's HF solutions:
\begin{eqnarray}
E^{\rm def}_{\rm HF}&=&{1 \over 2} \left(\sum_{i'\in{\rm
                           frag}}\left[2h^{\rm def}_{i'i'}+2F^{\rm
    def}_{i'i'}\right]\right)+E^{\rm def}_{\rm nuc},\label{eq:defect_E}
  \end{eqnarray}                         
with the effective $E^{\rm def}_{\rm nuc}$:
\begin{widetext}
\begin{eqnarray}
  E^{\rm def}_{\rm nuc}&=&{1 \over 2}\sum_{K'\in{\rm frag}}\sum_{L'\in{\rm frag}}\phantom{}^{'}{Z_{K'}Z_{L'} \over |{\bf R}_{K'}-{\bf
    R}_{L'}|}+\sum_{K'\in{\rm frag}} Z_{K'}\cdot V\left({\bf
                          R}_{K'}\right)\nonumber\\
  &&-2\sum_{i\in{\rm frag}}\left< i\left| -\sum_{K'\in{\rm frag}}{Z_{K'} \over |{\bf r}-{\bf  R}_{K'}|}\right|i\right>-\sum_{K\in{\rm frag}}\sum_{K'\in{\rm frag}}\phantom{}^{'}{Z_{K}Z_{K'} \over |{\bf R}_{K}-{\bf R}_{K'}|}.\label{eq:E_nuc_defect}
\end{eqnarray}
\end{widetext}
  The first term in (\ref{eq:E_nuc_defect}) is merely the energy of the clamped nuclei at the new position. The second term involves the complete electrostatic potential from the previous geometry from
both electrons and nuclei $V$ at the new positions
of the nuclei ${\bf R}_{K'}$. Note that the potential $V$ implicitly includes the contributions from the fragment's
nuclei (at the old positions) and from the fragment electrons frozen in the periodic solution. These contributions are parasitic, as the correct energy due to the mutual interaction between the fragment nuclei and between the fragment nuclei and fragment electrons -- all after the defect formation -- are already taken into account (the first term of eq. (\ref{eq:E_nuc_defect}) and the first term of eq. (\ref{eq:defect_E}), respectively). Therefore these parasitic contributions have to be explicitly removed, which is done in the fourth and third terms of (\ref{eq:E_nuc_defect}), respectively.

The second, third, and fourth terms constitute the electrostatic potential from the frozen environment. In the case when the entire system is encompassed by the
fragment and there is no environment, the three will sum to zero and the
$E^{\rm def}_{\rm HF}$ will take the usual expression of molecular RHF. In case of unmodified embedded fragment, it instead reduces to
$E^{\rm  frozen\, frag}_{\rm HF}$ of eq. (\ref{eq:frozenfrag_E}).

Formally, the summations over $K$ and $K'$ are performed over all the fragment nuclei before and after formation of the defect,
respectively. However, it is neither necessary to include the unaltered fragment
nuclei (i.e. when $K$=$K'$) in (\ref{eq:E_nuc_defect}). As demonstrated in section \ref{sect:intens_nuc}, energy differences, which in fact are the physically relevant quantities, are immune to inclusion or exclusion of these atoms, as they add only a constant shift to the total HF energy of the aperiodic defect.

\section{Implementation of the aperiodic-defect model}\label{sec:implement}

The current implementation of the embedded aperiodic fragment involves a
consecutive run of several programs. First, a primitive unit-cell periodic HF
calculation, and localization of the occupied orbitals are done via the Crystal
code \cite{Erba2022}. In these calculations the new positions of the atoms of the
defect are occupied by ghost "placeholders" that contain no charge and an ultra-narrow
s-AO with a large Gaussian exponent (1,000,000 au) such that no electron can actually occupy it. In case of the ghost and initial
atom coinciding (e.g. in substitution defects), the former
is shifted by a very small distance (e.g. 0.00000001 {\AA}) with respect to the latter, as
otherwise the HF
calculation cannot be carried out.
Crystal also evaluates the
electrostatic potential $V\left({\bf R}_{K'}\right)$ at the position of the
manipulated atoms $K'$  to be used in
expression (7) of the main text.

Next, in a second, single-iteration Crystal run, the actual
AOs are added on the placeholder centers in order to evaluate the Fock matrix
in the basis of these AOs. Technically this is performed as a dual-basis set
 option of the Crystal code. 

This information is then transferred to the Cryscor code \cite{usvyat18}, which is
used to define the fragment in terms of the following quantities:
\begin{itemize}
\item The initial occupied orbitals of the fragment: $\{i\in {\rm frag}\}$
\item Atoms that are removed in the defect formation: $\{K\in {\rm frag}\}$
\item Atoms that are added in the defect formation: $\{K'\in {\rm frag}\}$
\item The atoms that serve as centers for the fragment's AOs: $\{\mu'\}$ 
  \end{itemize}  
 With this the fragment AOs $\mu'$ are constructed according to eq. (3) of the
 main text.
  
The next step in the calculation is evaluation of the 2- and 3-index integrals:
$(P|Q)$, $(\mu'\nu'|P)$, $(i\mu'|P)$  and $(ii|P)$, where the indices $P$ and $Q$ denote the fitting
functions. For evaluation of these integrals the periodic local density fitting
machinery of the Cryscor code \cite{Pisani2012} is used, as described in Ref. \onlinecite{usvyat2013}. The fit domain is chosen to be universal, coinciding with the atomic fragment for the $\mu'$ AOs, such that the
one-term robust fitting can be employed.

\begin{figure*}[h]
    \centering
    \includegraphics[width=1.0\textwidth]{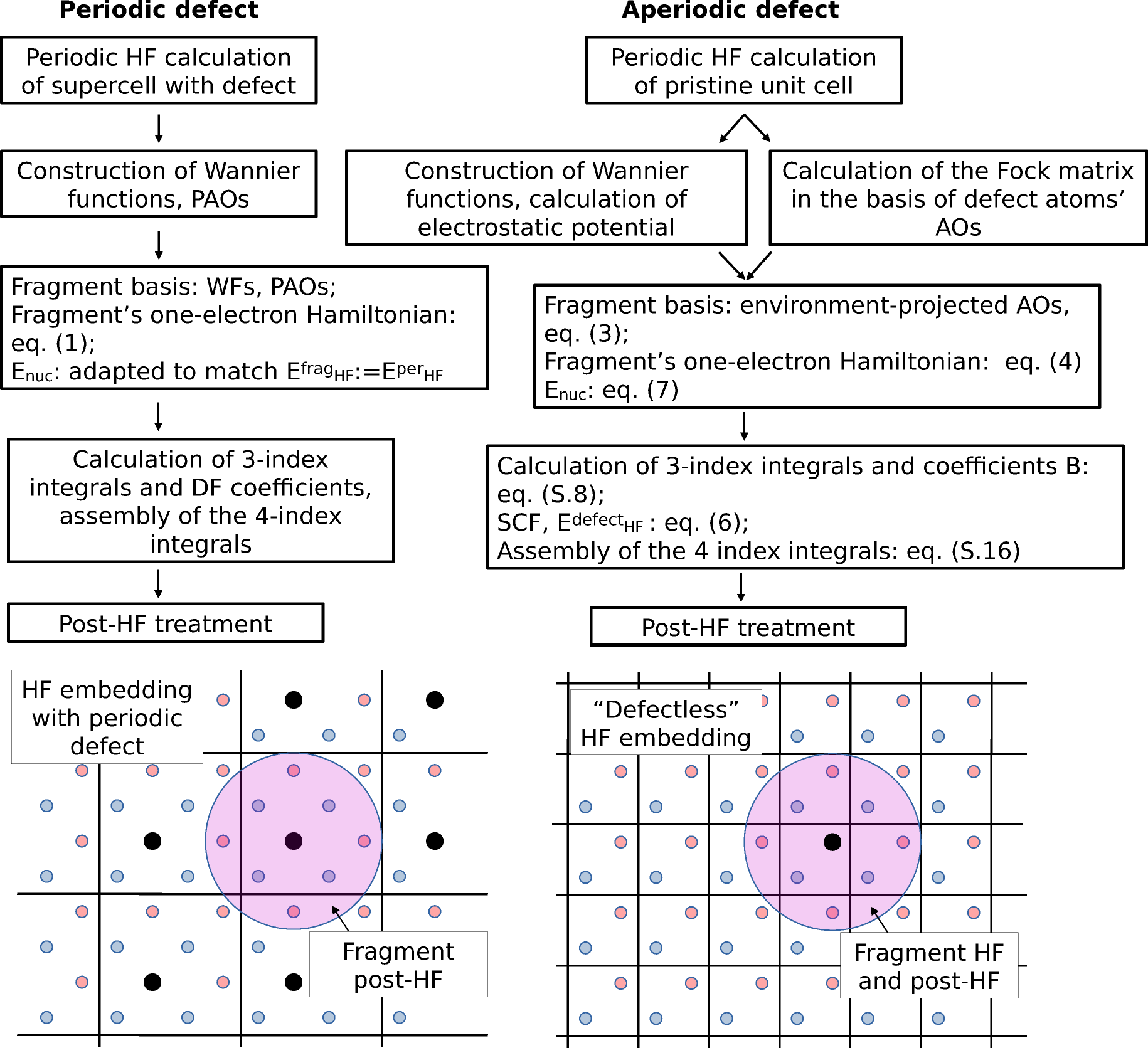}
    \caption{Flowcharts of the workflows for the embedded periodic defect of Ref. \onlinecite{christlmaier21} (left) and the embedded aperiodic defect presented here (right).}
    \label{fig:workflow}
\end{figure*}

Next, the  3-index intermediates $B_{\mu'\nu'}^{P}$ are calculated:
\begin{eqnarray}
  B_{\mu'\nu'}^{P}&=&\sum_{Q}\left(\mu'\nu'|Q\right)\left[J^{-1/2}\right]_{QP},\label{eq:B_coef}
 \end{eqnarray}                    
 with $J_{PQ}=(P|Q)$, as well as quantities $B^P_{i\nu'}$ and
$B^P_{ii}$, which are defined analogously to $B_{\mu'\nu'}^{P}$.

These intermediates are first used to calculate the Coulomb and exchange
matrix elements subtracted from the periodic Fock matrix to evaluate the
fragment's one-electron Hamiltonian $h^{\rm frozen\, frag}$ (eq. (1) of the
main text) in the basis $\mu'$:
\begin{eqnarray}
 J_{\mu'\nu'}&=&\sum_{P} B_{\mu'\nu'}^{P}\sum_{i}B_{ii}^{P},\label{eq:frag_coul1} 
\end{eqnarray}
\begin{eqnarray}
  K_{\mu',\nu'}&=&\sum_{Pi}B_{i\mu'}^{P}B_{i\nu'}^{P}.\label{eq:frag_exch1} 
\end{eqnarray}
Further, adding the one-electron integrals to $h^{\rm frozen\,
  frag}_{\mu'\nu'}$ yields $h^{\rm def}_{\mu'\nu'}$ according to eq. (4) of
the main text.

In the SCF cycles the quantity $B_{\mu'\nu'}^{P}$ is used for the actual defective fragment's Coulomb and exchange contributions  
\begin{eqnarray}
 J^{\rm def}_{\mu'\nu'}&=&\sum_{P} B_{\mu'\nu'}^{P}\sum_{\rho'\sigma'}B_{\rho'\sigma'}^{P}D_{\rho'\sigma'}\label{eq:frag_coul} 
\end{eqnarray}
\begin{eqnarray}
  K^{\rm def}_{\mu',\nu'}&=&\sum_{P}B_{i'\mu'}^{P}B_{i'\nu'}^{P}\label{eq:frag_exch} 
\end{eqnarray}
to the fragment Fock matrix
\begin{eqnarray}\label{eq:frag_fock} 
  F^{\rm def}_{\mu'\nu'}&=&h^{\rm def}_{\mu'\nu'}+2J^{\rm def}_{\mu'\nu'}-K^{\rm def}_{\mu',\nu'}.
\end{eqnarray}
Here $D_{\rho'\sigma'}$ is the density matrix
\begin{eqnarray}
  D_{\mu'\nu'}&=2&\sum_{i'}C_{\mu' i'}C_{\nu' i'},\label{eq:dens} 
\end{eqnarray}
$B_{i'\nu'}^{P}$ is the half transformed intermediate
\begin{eqnarray}
  B_{i'\nu'}^{P}&=&\sum_{\mu'}B_{\mu'\nu'}^{P}C_{\mu' i'},\label{eq:half_tr} 
\end{eqnarray}
and $C_{\mu'i'}$ are the orbital expansion coefficients in the fragment basis. 
The HF energy is evaluated via the expressions (6) and (7) of the main text. The SCF is accelerated using direct inversion of the iterative subspace (DIIS) \cite{Pulay1980}.

After the SCF has converged, one can readily get to canonical post-HF treatment via
the FCIDUMP \cite{Knowles89} interface. For this, we assemble the 4-index integrals from
the 3-index quantities $B^P_{\mu'\nu'}$ of eq. (\ref{eq:B_coef})
transformed to the basis of active orbitals $r'$, $s'$, ... .
\begin{eqnarray}
  \left(r's'|t'u'\right)=\sum_{P}B_{r's'}^{P}B_{t'u'}^{P}.\label{eq:4ind} 
\end{eqnarray}

Figure \ref{fig:workflow} outlines this workflow and how it compares to that of the embedding scheme of Ref. \onlinecite{christlmaier21} (denoted as periodic defect).

\section{Size-intensivity of the energy differences}\label{sec:intens}
Although expression (11) of the main text for the fragment HF energy is
not strictly size-extensive, it possesses a more important quantity --
asymptotic size-intensivity for the energy differences. Below, we demonstrate
this.

\subsection{Fragment nuclei}\label{sect:intens_nuc}

Firstly we focus on the nuclei. Consider two systems A and B that deviate from
each other by the position and/or type of some atoms in the fragment. For
example A and B could be the system with and without defect. The
fragment nuclei that are the same in both A and B we denote as ``fixed''
($K'\in {\rm fixed}$),
while the ones that differ in A and B we mark by the corresponding system index: $K_A'\in A$
and $K_B' \in B$.
The energy difference $\Delta E^{\rm def}_{\rm HF}=E^{\rm def}_{\rm
  HF}(A)-E^{\rm def}_{\rm HF}(B)$ will then be:
\begin{widetext}
\begin{eqnarray}
\Delta E^{\rm def}_{\rm HF}&=&2\sum_{i_A'\in{\rm frag}}\left< i_A'\left| -{1\over 2} \nabla^2\right| i_A'\right>-2\sum_{i_B'\in{\rm frag}}\left< i_B'\left| -{1\over 2} \nabla^2\right| i_B'\right>\nonumber\\
                        && - 2\sum_{i_A'\in{\rm frag}}
                         \left[
 \left< i_A'\left| \sum_{K'\in{\rm fixed}}{Z_{K'} \over |{\bf r}-{\bf R}_{K'}|}\right|i_A'\right>+
  \left< i_A'\left| \sum_{K\notin{\rm frag}}{Z_{K} \over |{\bf r}-{\bf
                           R}_{K}|}\right|i_A'\right>\right]\nonumber\\
  &&+ 2\sum_{i_B'\in{\rm frag}}
                         \left[
 \left< i_B'\left| \sum_{K'\in{\rm fixed}}{Z_{K'} \over |{\bf r}-{\bf R}_{K'}|}\right|i_B'\right>+
  \left< i_B'\left| \sum_{K\notin{\rm frag}}{Z_{K} \over |{\bf r}-{\bf
     R}_{K}|}\right|i_B'\right>\right]\nonumber\\
    &&- 2\sum_{i_A'\in{\rm frag}}                         
 \left< i_A'\left| \sum_{K_A'\in A}{Z_{K_A'} \over |{\bf r}-{\bf R}_{K_A'}|}\right|i_A'\right>+
  2\sum_{i_B'\in{\rm frag}}\left< i_B'\left| \sum_{K_B'\in B}{Z_{K_B'} \over |{\bf r}-{\bf
                           R}_{K_B'}|}\right|i_B'\right>\nonumber\\ 
&&+ \sum_{i_A'\in{\rm frag}}\sum_{j\notin{\rm frag}}\left[4\left(i_A'i_A'|jj\right)-2\left(i_A'j|ji_A'\right)\right] - \sum_{i_B'\in{\rm frag}}\sum_{j\notin{\rm frag}}\left[4\left(i_B'i_B'|jj\right)-2\left(i_B'j|ji_B'\right)\right]                       
                         \nonumber\\
                    &&
+{1 \over 2}\sum_{i_A'\in{\rm frag}}\sum_{j_A'\in{\rm frag}}\left[4\left(i_A'i_A'|j_A'j_A'\right)-2\left(i_A'j_A'|j_A'i_A'\right)\right]\nonumber\\&&-{1 \over 2}\sum_{i_B'\in{\rm frag}}\sum_{j_B'\in{\rm frag}}\left[4\left(i_B'i_B'|j_B'j_B'\right)-2\left(i_B'j_B'|j_B'i_B'\right)\right]\nonumber\\
&&+\sum_{L'\in{\rm fixed}}\sum_{K_A'\in A}\phantom{}^{'}{Z_{L'}Z_{K_A'} \over |{\bf R}_{L'}-{\bf
    R}_{K_A'}|}-\sum_{L'\in{\rm fixed}}\sum_{K_B'\in B}\phantom{}^{'}{Z_{L'}Z_{K_B'} \over |{\bf R}_{L'}-{\bf
   R}_{K_B'}|}\nonumber\\
&&+{1 \over 2}\sum_{K_A'\in A}\sum_{L_A'\in A}\phantom{}^{'}{Z_{K_A'}Z_{L_A'} \over |{\bf R}_{K_A'}-{\bf
    R}_{L_A'}|}-{1 \over 2}\sum_{K_B'\in B}\sum_{L_B'\in B}\phantom{}^{'}{Z_{K_B'}Z_{L_B'} \over |{\bf R}_{K_B'}-{\bf
    R}_{L_B'}|}\nonumber\\  
&&- 2\sum_{i\notin{\rm frag}}\left< i\left| \sum_{K_A'\in A}{Z_{K_A'} \over |{\bf r}-{\bf
   R}_{K_A'}|}\right|i\right>+ 2\sum_{i\notin{\rm frag}}\left< i\left|
   \sum_{K_B'\in B}{Z_{K_B'} \over |{\bf r}-{\bf
   R}_{K_B'}|}\right|i\right>\nonumber\\
  && +\sum_{L\notin{\rm frag}}\sum_{K_A'\in A}{Z_{L}Z_{K_A'} \over |{\bf R}_L-{\bf
   R}_{K_A'}|}-\sum_{L\notin{\rm frag}}\sum_{K_B'\in B}{Z_{L}Z_{K_B'} \over |{\bf R}_L-{\bf
   R}_{K_B'}|}.\label{eq:E_HF_fixed}
\end{eqnarray}
\end{widetext}
The fixed atoms ($K'\in \rm fixed$ or $L'\in \rm fixed$) and environment atoms  ($K\notin \rm frag$ or $L\notin \rm frag$) appear in the same expressions: terms 3 and 4, terms 5 and 6, terms 13 and 19, and terms 14 and 20. This shows that $\Delta E^{\rm def}_{\rm HF}$ does not depend on whether the
``fixed'' nuclei are included in
the fragment or in the environment. Even though an inclusion of a nucleus in a
fragment does generally affect the fragments total HF energy, the energy difference will
remain the same unless this nuclei has a different relative position/type in the
respective systems. In other words the energy difference $\Delta E^{\rm
  frag}_{\rm HF}$ is size-intensive with respect to expansion of the nuclei set beyond
the explicitly manipulated ones.

Therefore in practical calculations, the summations over the fragment nuclei
$K$ and $K'$ in eqs.  (4) and
(7) of the main text do not need to be performed over the ``fixed''
ones. 

\smallskip

\subsection{Fragment electrons} 

Now let us assume that our fragment is large enough that a certain part of the
fragment's localized occupied orbitals at the boundary of the fragment does
not feel the presence of the defect and remain the same as in the initial bulk
calculation, we will denote such orbitals as ``bulk'': $i'\in {\rm
  bulk}$, in contrast to the orbitals $i_A'\in {\rm frag}$ and $i_B'\in {\rm frag}$. Then the energy difference $\Delta E^{\rm def}_{\rm HF}$ will take
the form:
\begin{widetext}
\begin{eqnarray}
\Delta E^{\rm def}_{\rm HF}&=&2\sum_{i_A'\in{\rm frag}}\left< i_A'\left| -{1\over 2} \nabla^2\right| i_A'\right>-2\sum_{i_B'\in{\rm frag}}\left< i_B'\left| -{1\over 2} \nabla^2\right| i_B'\right>\nonumber\\
                        && - 2\sum_{i_A'\in{\rm frag}}
  \left< i_A'\left| \sum_{K\notin{\rm frag}}{Z_{K} \over |{\bf r}-{\bf
                           R}_{K}|}\right|i_A'\right>+ 2\sum_{i_B'\in{\rm frag}}
  \left< i_B'\left| \sum_{K\notin{\rm frag}}{Z_{K} \over |{\bf r}-{\bf
     R}_{K}|}\right|i_B'\right>\nonumber\\
    &&- 2\sum_{i_A'\in{\rm frag}}                         
 \left< i_A'\left| \sum_{K_A'\in A}{Z_{K_A'} \over |{\bf r}-{\bf R}_{K_A'}|}\right|i_A'\right>+
  2\sum_{i_B'\in{\rm frag}}\left< i_B'\left| \sum_{K_B'\in B}{Z_{K_B'} \over |{\bf r}-{\bf
       R}_{K_B'}|}\right|i_B'\right>\nonumber\\
  &&- 2\sum_{i'\in{\rm bulk}}                         
 \left< i'\left| \sum_{K_A'\in A}{Z_{K_A'} \over |{\bf r}-{\bf R}_{K_A'}|}\right|i'\right>+
   2\sum_{i'\in{\rm bulk}}\left< i'\left| \sum_{K_B'\in B}{Z_{K_B'} \over |{\bf r}-{\bf
                           R}_{K_B'}|}\right|i'\right>\nonumber\\ 
&&+ \sum_{i_A'\in{\rm frag}}\sum_{j\notin{\rm frag}}\left[4\left(i_A'i_A'|jj\right)-2\left(i_A'j|ji_A'\right)\right] - \sum_{i_B'\in{\rm frag}}\sum_{j\notin{\rm frag}}\left[4\left(i_B'i_B'|jj\right)-2\left(i_B'j|ji_B'\right)\right]                       
                         \nonumber\\
                    &&
+{1 \over 2}\sum_{i_A'\in{\rm frag}}\sum_{j_A'\in{\rm
                       frag}}\left[4\left(i_A'i_A'|j_A'j_A'\right)-2\left(i_A'j_A'|j_A'i_A'\right)\right]\nonumber\\&&-{1
                       \over 2}\sum_{i_B'\in{\rm frag}}\sum_{j_B'\in{\rm
                       frag}}\left[4\left(i_B'i_B'|j_B'j_B'\right)-2\left(i_B'j_B'|j_B'i_B'\right)\right]\nonumber\\
   &&
+\sum_{i_A'\in{\rm frag}}\sum_{j'\in{\rm bulk}}\left[4\left(i_A'i_A'|j'j'\right)-2\left(i_A'j'|j'i_A'\right)\right]-\sum_{i_B'\in{\rm frag}}\sum_{j'\in{\rm bulk}}\left[4\left(i_B'i_B'|j'j'\right)-2\left(i_B'j'|j'i_B'\right)\right]\nonumber\\
&&+{1 \over 2}\sum_{K_A'\in A}\sum_{L_A'\in A}\phantom{}^{'}{Z_{K_A'}Z_{L_A'} \over |{\bf R}_{K_A'}-{\bf
    R}_{L_A'}|}-{1 \over 2}\sum_{K_B'\in B}\sum_{L_B'\in B}\phantom{}^{'}{Z_{K_B'}Z_{L_B'} \over |{\bf R}_{K_B'}-{\bf
    R}_{L_B'}|}\nonumber\\  
&&- 2\sum_{i\notin{\rm frag}}\left< i\left| \sum_{K_A'\in A}{Z_{K_A'} \over |{\bf r}-{\bf
   R}_{K_A'}|}\right|i\right>+ 2\sum_{i\notin{\rm frag}}\left< i\left|
   \sum_{K_B'\in B}{Z_{K_B'} \over |{\bf r}-{\bf
   R}_{K_B'}|}\right|i\right>\nonumber\\
  && +\sum_{L\notin{\rm frag}}\sum_{K_A'\in A}{Z_{L}Z_{K_A'} \over |{\bf R}_L-{\bf
   R}_{K_A'}|}-\sum_{L\notin{\rm frag}}\sum_{K_B'\in B}{Z_{L}Z_{K_B'} \over |{\bf R}_L-{\bf
   R}_{K_B'}|}.\label{eq:E_HF_bulk}
\end{eqnarray}
\end{widetext}
Again  $\Delta E^{\rm frag}_{\rm HF}$ becomes insensitive to inclusion of the
bulk region of the fragment into environment (compare the terms 7 and 17, terms 8 and 18, terms 13 and 9, and terms 14 and 10). It shows that asymptotically $\Delta E^{\rm frag}_{\rm HF}$ is size-intensive with respect to expansion of the
fragment and its value converges to the TDL.

\section{Computational details: Bond dissociation in Fluorographane}

\begin{figure}[H]
    \centering
    \includegraphics[width=0.45\textwidth]{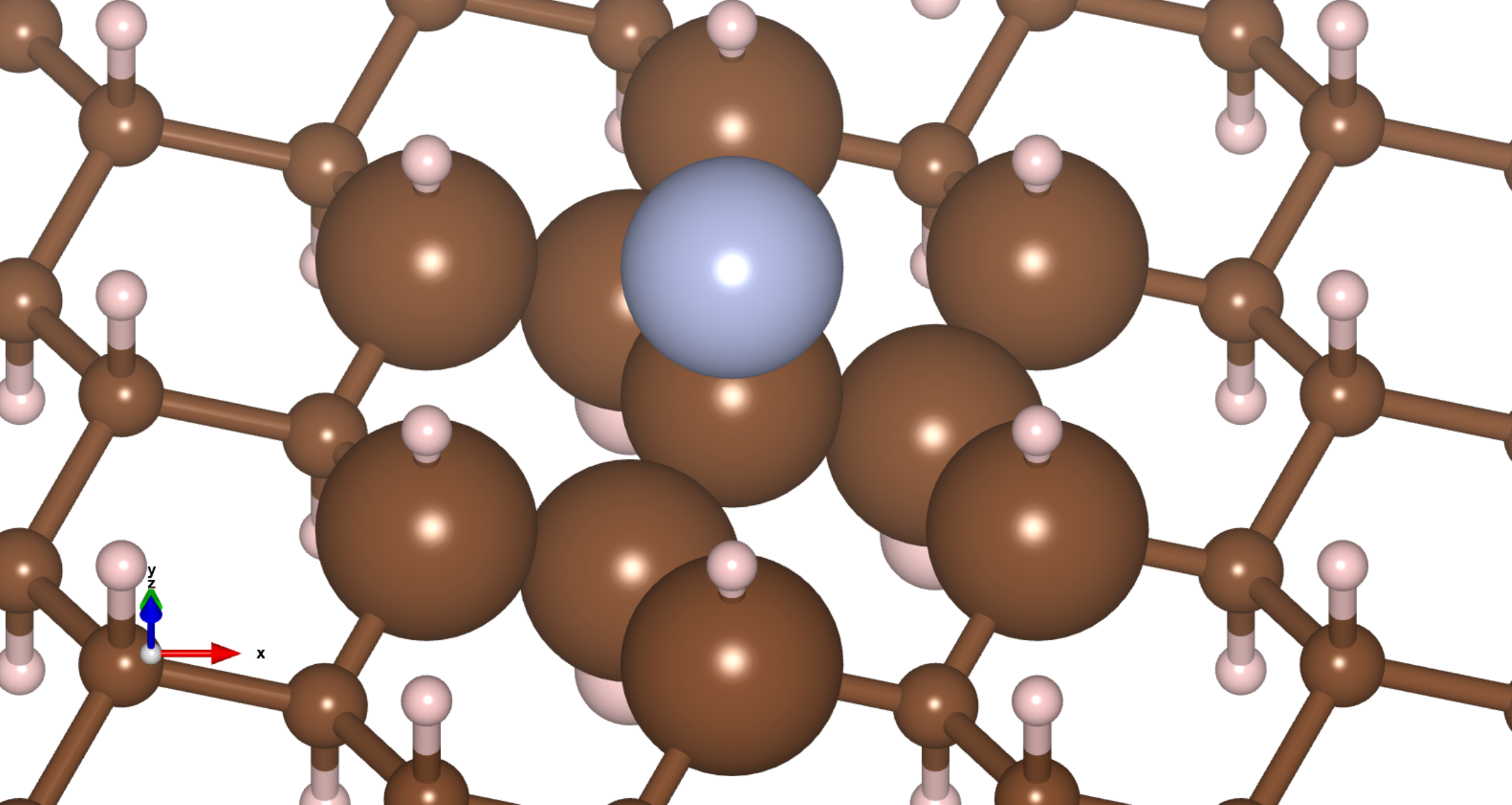}
    \caption{The H$\rightarrow$F substitution defect in graphane at its equilibrium bond length and, as an example, a 14-atom fragment for embedded-fragment calculations. The fragment is shown as space filling (noting that three hydrogen atoms on the underside of the monolayer are included), while the environment is shown as "ball-and-stick." Carbon atoms are brown, hydrogen atoms are pink, and the fluorine atom is grey.}
    \label{fig:fluorographane}
\end{figure}

As a test system we use a 2D graphane sheet, with a defect being a fluorine atom substituting one of the hydrogen atoms. We refer to this system as fluorographane. The aperiodic defect model of fluorographane with a possible choice of fragment is shown in Figure \ref{fig:fluorographane}.

The initial structure of graphane was optimized at the periodic B3LYP-D3 level \cite{Lee1988,Grimme2010} with the pob-TZVP-rev2 basis \cite{VilelaOliveira2019}.
For the aperiodic model we performed single-point periodic HF calculations on this pristine graphane with primitive unit cell with the same basis set. For technical reasons, see section \ref{sec:implement}, for each dissociation geometry the placeholder ghost atoms had to be added at the new positions of the carbon and fluorine atoms of the defect C-F bond.
For the minimum geometry of the defect, the positions of the fluorine atom and the corresponding carbon atom were optimized also with periodic B3LYP-D3 on a 3 by 3 supercell with all other atoms frozen. For each C-F bond length considered the position of the carbon atom was not re-optimized. In these optimization and further in the fragment calculations calculations also the pob-TZVP-rev2 basis was used.

Formation of the defect implied a removal of carbon and a hydrogen atom and addition of carbon (at the new position) and a fluorine atom (corresponding to the varied bond length).
After the fragment's HF within the aperiodic defect model or periodic HF within the periodic defect model, we performed post-HF calculations for different fragments via the FCIDUMP interface. For the density fitting, we used the fitting basis set optimized for MP2/aug-cc-pVTZ calculations \cite{weigend02b}.

\section{The raw data for the HF and CASPT2 calculations}

Table \ref{tab:raw_HF} contains the HF total energies for the minimum and bond-elongated structures of the fluorographane, as well as the the corresponding energy differences used in Fig. 2 of the main text. The CASPT2 total and dissociation energies  (only for $\Delta R=2$, see Fig.3 and the corresponding discussion in the main text) are compiled in table \ref{tab:raw_CASPT2}.

\begin{table*}[!ht]
  \centering
  \caption{The total and dissociation HF energies for the fluorine atom dissociation in fluorographane, calculated at 2{\AA} and 4{\AA} of the C-F bond elongation. $N_{\rm at}$ denotes the number of atoms in the fragment for the aperiodic defect model and the number in the supercell in the periodic defect model.}    
    \begin{tabular}{ccccccccccc}
    \hline\hline
    $N_{\rm at}$&~~ &$E(\Delta R=0)$, E$_{\rm h}$&~~ &   $E(\Delta R=2\AA)$, E$_{\rm h}$&~~ &  $E(\Delta R=4\AA)$, E$_{\rm h}$&~~ & $\Delta E(\Delta R=2\AA)$, eV& ~~ &  $\Delta E(\Delta R=4\AA)$, eV\\
    \hline
    \multicolumn{6}{l}{ Aperiodic defect model}\\
    5&&	-228.197997&&	-227.914305&&	-227.842800&&	7.720&&	9.665\\
    8&&	-254.214570&&	-253.980936&&	-253.924839&&	6.358&& 7.884\\
   14&&	-569.392963&&	-569.176844&&	-569.123728&&	5.881&&	7.326\\
   20&&	-684.282072&&	-684.075177&&	-684.019739&&	5.630&&	7.138\\
   23&&	-987.585908&&	-987.381216&&	-987.326695&&	5.570&&	7.054\\
   29&& -1470.730585&& -1470.527772&&   -1470.474211&&	5.519&&	6.976\\
   32&& -1573.194831&& -1572.993016&&   -1572.939620&&	5.492&&	6.945\\
   38&& -1764.169670&& -1763.969663&&   -1763.917001&&	5.442&&	6.875\\
   50&& -2713.733153&& -2713.534594&&   -2713.483379&&	5.403&&	6.797\\
    \hline
    \multicolumn{6}{l}{   Periodic defect model}\\
   16&&	-406.583853&&	-406.289814&&	-406.194613&&	8.001&&	10.592\\
   36&&	-791.196426&&	-790.956102&&	-790.847898&&	6.540&&	9.484\\
   64&& -1329.651688&&  -1329.435306&&  -1329.343168&&	5.888&&	8.395\\
  100&& -2021.950888&&  -2021.743738&&  -2021.670704&&	5.637&&	7.624\\
  144&& -2868.094247&&  -2867.891264&&  -2867.828459&&	5.523&&	7.232\\
  196&& -3868.081814&&  -3867.880882&&  -3867.823446&&	5.468&&	7.030\\
  256&& -5021.913605&&  -5021.713799&&  -5021.659398&&	5.437&&	6.917\\
  324&& -6329.589626&&  -6329.390483&&  -6329.337913&&	5.419&&	6.849\\
    \hline \hline
    \end{tabular}
    \label{tab:raw_HF}
\end{table*}

\begin{table*}[!ht]
  \centering
  \caption{The total and dissociation CASPT2(6,7) energies for the fluorine atom dissociation in fluorographane, calculated at 2{\AA} of the C-F bond elongation.}    
    \begin{tabular}{ccccccccc}
    \hline\hline
    $N_{\rm at}$(supercell)&~~ & $N_{\rm at}$(fragment)&~~ & $E(\Delta R=0)$, E$_{\rm h}$&~~ &   $E(\Delta R=2\AA)$, E$_{\rm h}$&~~ &  $\Delta E(\Delta R=2\AA)$, eV\\
    \hline
    \multicolumn{5}{l}{   Aperiodic defect model}\\
   &&	8 &&	-254.660410&&	-254.479438&&	4.924\\
   &&	14&&	-570.163203&&	-569.973301&&	5.167\\
   &&	20&&	-685.289361&&	-685.098846&&	5.184\\
   &&	23&&	-988.917887&&	-988.728013&&	5.167\\
    \hline
    \multicolumn{5}{l}{    Periodic defect model}\\
16 &&	8 &&	-407.030521&&	-406.881871&&	4.045\\
   &&	14&&	-407.355411&&	-407.199071&&	4.254\\
   &&	20&&	-407.593126&&	-407.438838&&	4.198\\
   &&	23&&	-407.920961&&	-407.766169&&	4.212\\
64 &&	8 &&	-1330.107763&&	-1329.898361&&	5.698\\
   &&	14&&	-1330.422863&&	-1330.231799&&	5.199\\
   &&	20&&	-1330.660704&&	-1330.476051&&	5.025\\
   &&	23&&	-1330.986137&&	-1330.804232&&	4.950\\
144&&	8 &&	-2868.552704&&	-2868.323690&&	6.232\\
   &&	14&&	-2868.865385&&	-2868.664331&&	5.471\\
   &&	20&&	-2869.103237&&	-2868.907809&&	5.318\\
   &&	23&&	-2869.428654&&	-2869.235830&&	5.247\\
    \hline \hline
    \end{tabular}
    \label{tab:raw_CASPT2}
\end{table*}

\newpage

\section{The TDL extrapolation of the HF dissociation energy}

An estimate of the TDL values can generally be achieved via extrapolation. Here we analyze the asymptotic decay of the finite-size error in periodic and
aperiodic defect models for the HF dissociation energy in Fluorographane.

In the periodic model the source if the finite size error is the unphysical
dipole-dipole interaction between the defect images. A large dipole moment
occurs in the dissociated geometry in the HF method, which favors a purely
ionic dissociation for this system (see Fig. 1 of the main text). With progressively and uniformly expanded
supercell, this erroneous interaction fades off as $1/R^3$,
where $R$ is distance between the defect images. In a 2D system, this distance is
proportional to the square root of the number of atoms in the supercell $N$,
suggesting an asymptotic $1/N^{3/2}$ convergence rate for the periodic model.

 In the aperiodic defect model, the leading error source is the unaccounted
 induction of the dipole moment of the defect into the environment. Pairwise
 dipole - induced dipole interaction decays as $1/R^6$. Here, however, one has
 to integrate this interaction over the 2D fragment boundary and from the
 boundary to infinity. These integrations slow down the asymptotic decay of
 the remaining finite size error to  $1/R^4$, where $R$ is a characteristic
 radius of the fragment. In a 2D system that corresponds to a $1/N^{2}$
 convergence rate, where $N$ is the number of atoms in the fragment.

\section{Potential energy curves for Fluorine atom dissociation from graphane}

\begin{figure}[H]
    \centering
    \includegraphics[width=0.5\textwidth]{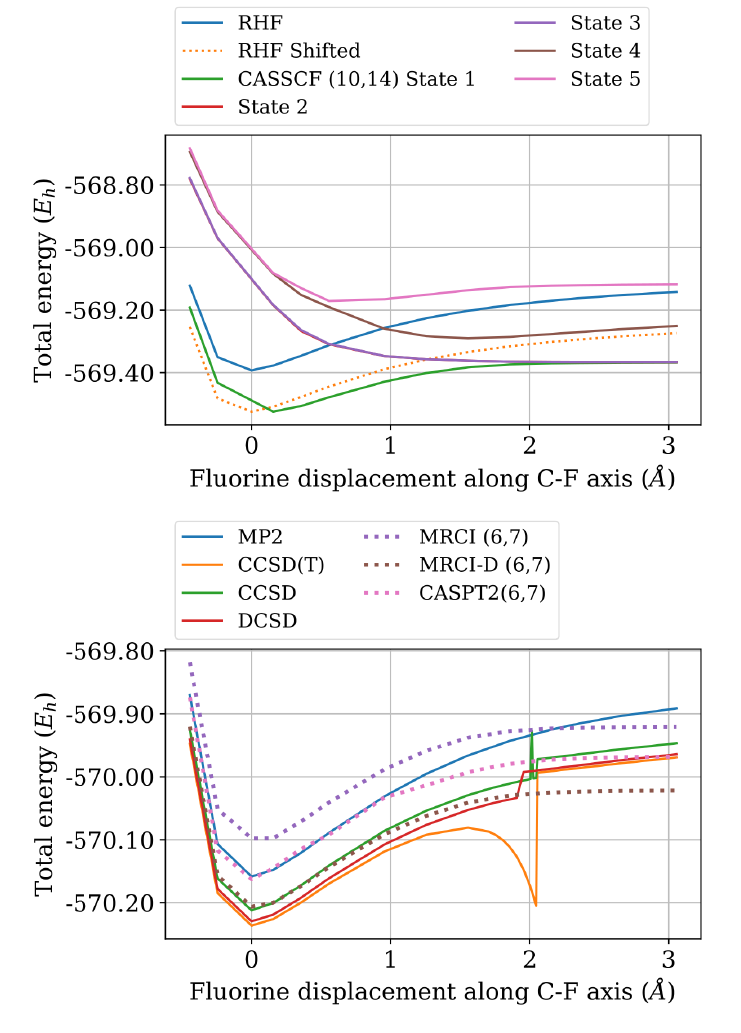}
    \caption{Potential energy curves for a single fluorine
      atom dissociation in a fluorine-hydrogen substitution defect in graphane
      using the aperiodic fragment approach and 14-atom fragment (see Fig. \ref{fig:fluorographane}). Top: RHF, energy-shifted RHF and
      SA(5)-CASSCF(10,14) with 5 lowest singlet states. States 2 and 3 are degenerate along the entire curve, and the avoided crossing occurs between states 1 and 4. Bottom: Methods which include dynamical correlation (the lowest state only): MRCI, MRCI
      with Davidson correction, CASPT2 (all with (6,7) active space), MP2,
      DCSD, CCSD and CCSD(T). Results of single-reference methods are shown with solid lines; those of multireference methods are shown with dashed lines.} 
    \label{fig:curve_panel}
\end{figure}

We start our analysis with the RHF and CASSCF results shown at the top
panel of Figure \ref{fig:curve_panel}. Neither of these methods include
dynamic correlation, but CASSCF can capture static correlation and also be
used for excited states. The RHF curve, although reasonable near the minimum,
sticks to the ionic dissociation, which is an excited state here. Qualitatively, it is similar to the periodic HF curve of the supercell
approach (Fig. 2 of Ref. \onlinecite{christlmaier21}), but less steep. The
reason for an additional steepness is the accumulated static correlation and electrostatics in modestly sized supercell used in that study, which add an energy penalty (see Figs. 1 and 2 of the main text).  

The CASSCF 
curves, despite the lack of dynamic correlation, describe the neutral
dissociation qualitatively correctly. Moreover, with CASSCF, the whole
landscape of the local excited states can be reconstructed. In this system, it
demonstrates the avoided crossing between states 1 and 4 and the switch of dissociative ground state character from ionic to neutral at about 1.5 {\AA} of the bond elongation. 

The bottom panel of Figure \ref{fig:curve_panel} compiles the methods that
include dynamic correlation. Firstly we note that the multireference
methods, MRCI \cite{10.1063/1.455556},
MRCI-D (i.e. MRCI with Davidson's fixed size-extensivity correction, sometimes referred to as MRCI+Q), and CASPT2\cite{Celani2000}, correct the CASSCF bond
length minimum, showing the importance of dynamic correlation. As
expected, they also correctly describe the neutral dissociation limit. The
single reference methods tested here all mutually agree  near the
minimum. All of them 
 reproduce
 the ionic dissociation, however. MP2
 follows HF and stays on the ionic
 state along the whole dissociation curve. Distinguishable cluster singles and
 doubles (DCSD) \cite{Kats2013} and CCSD
do switch to the neutral state at the avoided crossing point, but very soon
become unstable due to the growth of static correlation in this state.
This growth is evident
from the divergent behavior of the perturbative triples
contribution, (T). At some point (near 2 {\AA} of the bond elongation), all
these methods switch to the ionic state.

\begin{figure}[H]
    \centering
    \includegraphics[width=0.5\textwidth]{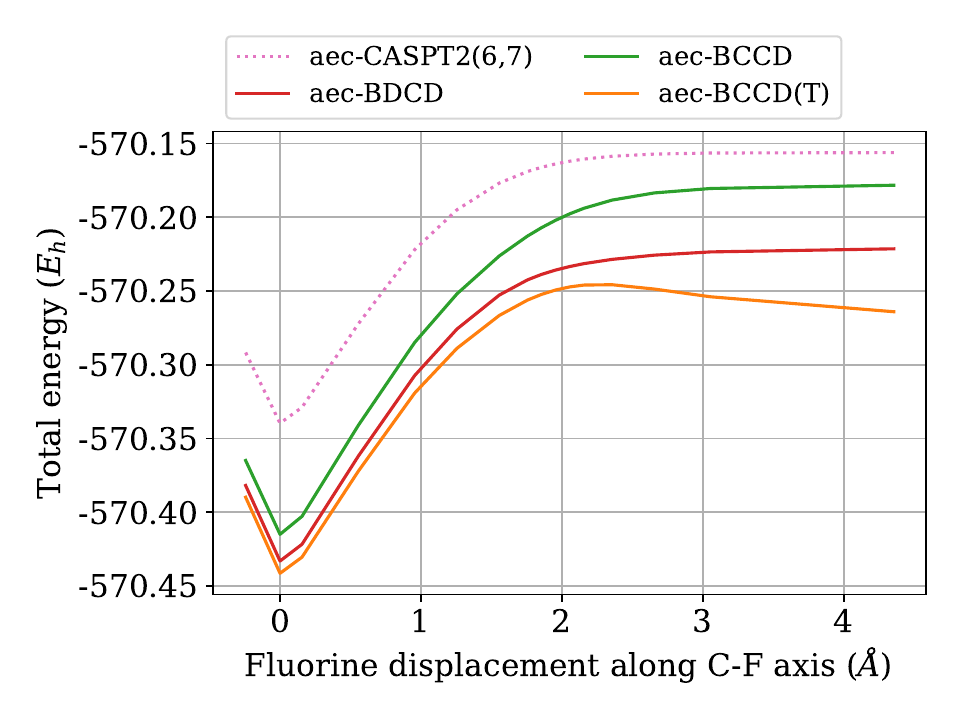}
    \caption{BDCD, BCCD, BCCD(T) and CASPT2 (6,7) potential energy curves for
      a single fluorine atom dissociation in a fluorine-hydrogen substitution
      defect in graphane using the aperiodic fragment approach and 14-atom fragment (see Fig. \ref{fig:fluorographane}). "aec" stands for
      "all-electron correlated": all Brueckner and CASSCF orbitals are
      re-optimized at each geometry with the orbitals from the previous
      geometry taken as starting guess, so the frozen-core approximation is not employed.}
    \label{fig:bccd_curve}
\end{figure}

The problem of instability of DCSD method at fluorine dissociation is somewhat curious,
as it is known to be capable of dealing with static correlation better than
most single-reference methods \cite{doi:10.1063/1.4940398}. In order to investigate this, in Figure \ref{fig:bccd_curve}, we
employ Brueckner distinguishable cluster doubles (BDCD) \cite{Kats2014}
along with Brueckner coupled cluster doubles (BCCD) \cite{Handy1989}
,  BCCD(T), and CASPT2 as a reference. To facilitate
stability, the starting guesses for Brueckner orbitals along the dissociation
path were taken from the previous geometries.
To keep the order of
orbitals from one point to another, the FCIDUMP was written directly in the (symmetrically orthogonalized) basis orbitals $\mu'$. In contrast to DCSD, BDCD is indeed very
stable along the whole dissociation and fairly well reproduces the CASPT2
reference with only a slight overestimation of the dissociation energy. BCCD does not break down either, but its dissociation energy is much
larger than CASPT2's. Finally, the (T) correction exhibits its trademark asymptotic divergence
in the presence of strong correlation.

\typeout{}
\bibliography{citations}